\documentclass[10pt,epsfig]{article}

\usepackage{enumerate}
\usepackage{float}
\usepackage{subfig}
\usepackage{color}
\usepackage[table]{xcolor}
\usepackage{slashed}
\usepackage{multirow}
\usepackage{cite}

\let\counterwithin\relax
\usepackage{chngcntr}
\usepackage{amssymb, amsmath,mathrsfs}

\usepackage{graphics}
\usepackage{graphicx}
\usepackage{epsf}
\usepackage{epsfig}
\usepackage{float}
\usepackage{makecell}
\usepackage{multirow}
\usepackage{color}
\usepackage{xcolor}
\usepackage{simplewick}
\usepackage{longtable}

\usepackage[utf8]{inputenc}
\usepackage{amsmath}
\usepackage{amssymb}
\usepackage{subfig}
\usepackage[normalem]{ulem}

\usepackage{mathtools}
\usepackage{amsmath}
\usepackage{diagbox}

\newcommand\undermat[2]{
	\makebox[0.5pt][l]{$\smash{\underbrace{\phantom{%
					\begin{matrix}#2\end{matrix}}}_{ \let\scriptstyle\textstyle\text{\large $#1$}}}$}#2}
\newcommand\overmat[2]{
	\makebox[-1pt][l]{$\smash{\overbrace{\phantom{%
					\begin{matrix}#2\end{matrix}}}^{ \let\scriptstyle\textstyle\text{\large $#1$}}}$}#2}    
\usepackage{tikz}
\usepackage[vcentermath]{youngtab}
\usepackage{slashed} 
\usepackage[font=small]{caption} 
\usepackage{times} 

\usepackage{bm}       
\usepackage{bbm} 

\usepackage{relsize}  

\usepackage{autobreak}  

\usepackage[colorlinks=true,
linkcolor=blue,
urlcolor=red,
citecolor=red]{hyperref}

\usepackage{makeidx} 
\usepackage{bbding} 
\usepackage{listings} 
\usepackage{ytableau}
\usepackage[utf8]{inputenc}
\usepackage[T1]{fontenc}
\usepackage{lmodern}

\ytableausetup
{boxsize=1.0em}

\long\def\rpl#1!!#2!!{\textcolor{red}{#1} \textcolor{blue}{#2}}

\usepackage[top=1in, left=0.95in, bottom=1.1in, right=0.95in]{geometry}
\def\baselinestretch{1.27}
\usepackage[toc]{appendix}

\newcommand{\Rd}[1] { {\color{red} #1}}
\newcommand{\Bl}[1] { {\color{blue} #1}}

\newcommand{\beq}{\begin {equation}}  
\newcommand{\eeq}{\end   {equation}} 
\newcommand{\bea}{\begin {eqnarray}} 
\newcommand{\eea}{\end   {eqnarray}}  
\newcommand{\baa}{\begin {array}   } 
\newcommand{\eaa}{\end   {array}   }     
\newcommand{\bit}{\begin {itemize} }
\newcommand{\eit}{\end   {itemize} }
\newcommand{\be }{\begin {equation}} 
\newcommand{\ee }{\end   {equation}}

\newcommand{\mc}[1]{\mathcal{#1}}

\newcommand{\vev}[1]{ \langle {#1} \rangle }

\newcommand{\hc}{{\text{h.c.}}~}

\newcommand{\eq}[1]{\begin{equation}\begin{split} #1 \end{split}\end{equation}}

\newcommand{\comment}[1]{}

\newcommand{\TeV}{\ensuremath{\mathrm{TeV}}}

\newcolumntype{M}[1]{>{\centering\arraybackslash}m{#1}}
\newcolumntype{N}{@{}m{0pt}@{}}

\allowdisplaybreaks

\begin{document}

\begin{center}


{\Large \textbf  {A Complete Set of the Dimension-8 Green's Basis Operators in the Standard Model Effective Field Theory}}\\[10mm]

Zhe Ren$^{a, b}$\footnote{renzhe@itp.ac.cn}, Jiang-Hao Yu$^{a, b, c, d, e}$\footnote{jhyu@itp.ac.cn}\\[10mm]

\noindent 
$^a${\em \small CAS Key Laboratory of Theoretical Physics, Institute of Theoretical Physics, Chinese Academy of Sciences,    \\ Beijing 100190, China}  \\
$^b${\em \small School of Physical Sciences, University of Chinese Academy of Sciences,   Beijing 100049, China}   \\
$^c${\em \small Center for High Energy Physics, Peking University, Beijing 100871, China} \\
$^d${\em \small School of Fundamental Physics and Mathematical Sciences, Hangzhou Institute for Advanced Study, UCAS, Hangzhou 310024, China} \\
$^e${\em \small International Centre for Theoretical Physics Asia-Pacific, Beijing/Hangzhou, China}\\[10mm]

\date{\today}   
          
\end{center}

\begin{abstract}
We present a complete and independent off-shell Green's basis of the dimension 8 operators in the Standard Model effective field theory (SMEFT). We propose an off-shell amplitude formalism such that this new kind of amplitudes has a one-to-one correspondence to general operators carrying all kinds of redundancies. The advantage of such formalism is that all kinds of redundancies can be formulated explicitly in terms of the off-shell spinors and thus be removed systematically from the exhaustive list of the general off-shell amplitudes. Following such procedure, redundant operators can be reduced to the ones with the only redundancy from the equation of motion kept, the Green's basis. We find there are (993+1649) independent (on-shell + off-shell) operators for one generation of fermions and (44807+66197) ones for three generations at the dimension 8. This systematic method can be applied to obtain the Green's basis at any mass dimension for general EFTs.

 %
 %
 %
 %
 %
 %
 %
 %
 %
 %
 %

\end{abstract}

\newpage

\setcounter{tocdepth}{4}
\setcounter{secnumdepth}{4}

\tableofcontents

\setcounter{footnote}{0}

\def\baselinestretch{1.5}
\counterwithin{equation}{section}

\newpage

\section{Introduction}

Being the most important achievement in particle physics, the Standard Model (SM) have been proved to be sufficient up to $\Lambda_{\rm NP} \sim \TeV$ scale by the experiments at the large hadron collider (LHC). Due to the energy gap between the $\Lambda_{\rm NP}$ scale and the energy scale of the SM particles, the physical effects of new physics particles below the $\Lambda_{\rm NP}$ scale can be characterised in the framework of effective field theory (EFT). More specifically, the Standard Model effective field theory (SMEFT) provides a systematical framework to parameterize all possible Lorentz-invariant new physics above the electroweak scale. Besides the SM Lagrangian, the Lagrangian of the SMEFT can be organized as a power expansion on canonical dimension in terms of the higher-dimensional operators composed of the SM field degrees of freedom (d.o.f.),
\bea
	{\mathcal L}_{\rm SMEFT} = {\mathcal L}_{\rm SM}^{(4)} + \sum_{d > 4 }\left( \frac{1}{\Lambda_{\rm NP}} \right)^{d-4} \sum_i C_i^{(d)} {\mathcal O}^{(d)}_{i},
\eea
where $C_i^{(d)}$ denotes the Wilson coefficient of operators $O_i^{(d)}$, of the canonical mass dimension $d$.

 At each mass dimension $d$, one can always write down a set of Lorentz and gauge invariant operators, however those operators might be redundant under the field re-definition (equivalently the equation of motion) and other equivalence relations, such as integration by part (IBP), covariant derivative commutator (CDC), the Schouten identity (equivalently the Fierz identity) and the Bianchi identity. Among these redundancies, the field re-definition plays a special role because operators related by the field re-definition give rise to the same S-matrix, and thus are physically equivalent~\cite{CHISHOLM1961469,Kamefuchi:1961sb}, which requires the fields in effective operators to be on-shell and equivalently means that operators related by the equation of motion (EOM) are redundant~\cite{Arzt:1993gz,Criado:2018sdb}. A set of operators with the EOM removed, are named as the on-shell operators, otherwise off-shell ones. Still other redundancies than the EOM needs to be removed. With all the redundancies eliminated, a complete and independent set of operators at the dimension $d$ forms {\it the on-shell physical basis}, while a complete and independent set of operators with only the EOM kept is called {\it the off-shell Green's basis}.

 According to the Weinberg's Folk theorem~\cite{Weinberg:1978kz}, for a generic EFT, with the d.o.f. and power counting identified, one should write all possible terms consistent with the global and gauge symmetries of the theory: the on-shell physical bases at each mass dimension. Pioneered by Weinberg, the on-shell physical bases at dimension 5, 6, 7, 8 and 9 have been listed in Ref.~\cite{Weinberg:1979sa,Buchmuller:1985jz,Grzadkowski:2010es,Lehman:2014jma,Liao:2016hru,Li:2020gnx,Murphy:2020rsh,Li:2020xlh,Liao:2020jmn}. Furthermore, with the light d.o.f., such as the sterile neutrinos, the operator basis has been constructed up to dimension 9~\cite{delAguila:2008ir,Aparici:2009fh,Bhattacharya:2015vja,Liao:2016qyd,Li:2021tsq}. Recently, a systematic and automated method~\cite{Li:2022tec} to construct on-shell physical bases for generic EFTs is introduced and applied to construct the on-shell physical bases at dimension 8 and 9 in the SMEFT~\cite{Li:2020gnx,Li:2020xlh} and on-shell physical bases in other EFTs~\cite{Li:2020tsi,Li:2021tsq, Sun:2022ssa, Sun:2022snw}.


On the other hand, the off-shell Green's basis is also quite useful, especially to deal with procedure of the matching and running between the UV new physics and the SMEFT~\cite{Jiang:2018pbd,Gherardi:2020det}. In the top-down approach~\footnote{There are two approaches: the top-down approach and the bottom-up approach, to relate the effective operators and their UV realizations. In Ref.~\cite{Li:2022abx}, a bottom-up approach is proposed where all the possible tree-level UV origins of the operators can be found by acting the Casimirs of the Poincare and gauge symmetries on the operators and obtained all possible UV resonances for the dimension-6 (also Ref.~\cite{deBlas:2017xtg}) and dimension-7 SMEFT operators, and dimension-8 bosonic operators. }, given a new physics model, integrating out all heavy particles gives rise to the operators that are generally related by the field re-definitions and other redundancies. There are typically two ways of performing the matching procedure: the diagrammatic approach and the path-integral approach. In the diagrammatic approach~\cite{Georgi:1993mps,Skiba:2010xn,Carmona:2021xtq}, one first calculates the one-light-particle irreducible (off-shell) Green's functions in the UV and then matches to the same Green's functions in the SMEFT, which is directly related to the off-shell Green's basis operators. Therefore, the operators obtained after integrating out the heavy particles should be identified as the Green's basis operators. On the other hand, in the path-integral approach~\cite{Gaillard:1985uh,Cheyette:1987qz,Henning:2014wua,Cohen:2020qvb,Fuentes-Martin:2020udw}, the effective action below the heavy particle scale can be expanded locally as the effective operators, and the Green's basis helps to reduce these operators into the on-shell physical basis by treating the EOM in the last step of the basis conversion.



 
 To obtain the Green's basis, one needs to keep operators related by the field re-definition, which is equivalent to keeping operators related by the EOM, while operators related by other redundant relations, including the IBP, the CDC, the Schouten identity and the Bianchi identity, should be considered redundant still. In the SMEFT, a dimension-6 Green's basis have been listed in Ref.~\cite{Gherardi:2020det}. A dimension-6 Green's basis in the SMEFT extended with sterile neutrinos~\cite{Chala:2020vqp} and a dimension-5 Green's basis in the SMEFT extended with an axion-like particle~\cite{Chala:2020wvs} have been listed as well. The bosonic sector of a dimension-8 Green's basis have been listed in Ref.~\cite{Chala:2021cgt}, yet the complete dimension-8 Green's basis have not been worked out due to the large number of independent operators and the redundancies in the flavor structures of operators involving fermions.

In this work, we propose a systematic method to construct the complete set of the Green's basis in the SMEFT, by extending the on-shell amplitude and operator correspondence, which lays out the foundation of the construction on the on-shell physical operator bases~\cite{Li:2020gnx,Li:2020xlh,Li:2022tec}. 
However, the Young tensor technique on the Lorentz structure of effective operator needs the on-shell amplitude carrying the $SL(2,C) \times SU(N)$ symmetry, which cannot be applied to the off-shell operators. 
In order to construct the off-shell operators, we propose an off-shell amplitude formalism so that any operators carrying various redundancies have one-to-one correspondences to the off-shell amplitudes expressed by the off-shell spinors. Thus the off-shell amplitudes are the ones with the EOM and the CDC inherited from the formalism of operators and it would recover the on-shell amplitudes by substituting the off-shell spinors with the on-shell spinors. In the off-shell amplitude formalism, one would first utilize the off-shell amplitudes to generate an over-complete set of operators by finding all possible contractions of off-shell spinors, and all the redundancy relations can be written explicitly in terms of the off-shell spinors, and applying these relations to the over-complete operator set one could remove any kinds of redundancies systematically and obtain the operators with only the EOM kept. Thus we are able to construct the Green's basis in the SMEFT through the above systematic method, which can also be applied to general EFTs to obtain the Green's basis at any mass dimension. The dimension-8 Green's basis in the SMEFT in our result includes 993+1649=2642 independent operators for one generation fermions, and 44807+66197=111004 independent operators for three generations of fermions. The statistic result of the counting is consistent with the counting using the Hilbert series~\cite{Henning:2015alf, Lehman:2015via, Henning:2017fpj} and Sym2Int~\cite{Fonseca:2019yya}. 

This paper is organized as follows. In section~\ref{sec:offshell}, we introduce the building blocks of the SMEFT, and propose the off-shell amplitude formalism, along with a one-to-one mapping from operators to the off-shell amplitudes, as a generalization of the amplitude-operator correspondence in Ref.~\cite{Li:2020gnx,Li:2020xlh,Li:2022tec}. In section~\ref{sec:Green}, we describe the method to perform the off-shell construction of the Green's basis with concrete examples. In section~\ref{sec:Operators}, we list a complete and independent dimension-8 Green's basis in the SMEFT, along with a statistic of the basis in table~\ref{tab:statistic1} and table~\ref{tab:statistic2}. We conclude in section~\ref{sec:con}.

\section{Building Blocks and Off-shell Amplitude Formalism}\label{sec:offshell}

In general EFTs, the building blocks of effective operators are fields and covariant derivatives. In this work, the fields and covariant derivatives in the SMEFT are presented as the irreducible representations of the $SL(2,\mathbb{C})$ group, that is,
\bea
&&\phi \in (0,0), \quad \psi_{\alpha }  \in (1/2,0), \quad \psi^{\dagger}_{\dot\alpha } \in (0,1/2), \\
&&F_{{\rm L}\alpha\beta} = \frac{i}{2}F_{\mu\nu}\sigma^{\mu\nu}_{\alpha\beta}\in (1,0),  \quad F_{{\rm R} \dot\alpha\dot\beta} = -\frac{i}{2}F_{\mu\nu}\bar\sigma^{\mu\nu}_{\dot\alpha\dot\beta}\in(0,1), \\
&&D_{\alpha\dot\alpha} = D_{\mu}\sigma^{\mu}_{\alpha\dot\alpha} \in (1/2,1/2).
\eea
where $F_{\rm L/R} = \frac{1}{2}(F \mp i\tilde{F})$ are gauge bosons in the chiral basis, with $\tilde{F}^{\mu \nu}=\frac{1}{2} \epsilon^{\mu \nu \rho \eta} F_{\rho \eta}$, $\epsilon^{0123}=1$, and $\psi$, $\psi^{\dagger}$ and $\phi$ denote the left-handed Weyl spinors, the right-handed Weyl spinors and scalars respectively. The covariant derivatives always act on the nearest field on the right, and derivative in this paper always means the covariant derivative unless otherwise specified.

With this notation, the following correspondence between the building blocks of operators and on-shell spinors has been introduced in Ref.~\cite{Li:2020gnx,Li:2020xlh},
\eq{
	\begin{array}{lcl}
		D^{r_i-1}F_{{\tiny\rm L/R}\,i}	&	\rightarrow	&	\lambda_i^{r_i\pm1}\tilde\lambda_i^{r_i\mp1}			\\
		D^{r_i-1/2}\psi^{(\dagger)}_i	&	\rightarrow	&	\lambda_i^{r_i\pm1/2}\tilde\lambda_i^{r_i\mp1/2}	\\
		D^{r_i}\phi_i	&	\rightarrow	&	\lambda_i^{r_i}\tilde\lambda_i^{r_i}
	\end{array},
}
where $i$ labels the $i$th field in an operator, $i \in \{1,2,\cdots,N\}$, and $N$ denotes the number of fields in the operator. $r_i$ is a positive integer or half-integer depending on the field is bosonic or fermionic. The correspondence can be recognized as a mapping from operators to on-shell amplitudes. It is obvious that the mapping is not one-to-one since the spinor indices of the fields and the spinor indices of the derivatives are not distinguished. For example, the corresponding amplitude form of both $D_{\alpha\dot{\alpha}} \psi_{i\beta}$ and $D_{\beta\dot{\alpha}} \psi_{i\alpha}$ is $\lambda_{i\alpha}\lambda_{i\beta}\tilde{\lambda}_{i\dot{\alpha}}$, and the difference between the two operator forms is
\begin{equation}\label{eq:egcorres}
    D_{\alpha\dot{\alpha}} \psi_{i\beta}-D_{\beta\dot{\alpha}} \psi_{i\alpha} = \epsilon_{\alpha\beta} D^{\gamma}{}_{\dot{\alpha}} \psi_{i\gamma},
\end{equation}
where in our notation
\begin{eqnarray}
    \epsilon^{\alpha\beta}=\left(\begin{array}{cc}
		0 & 1 \\
		-1 & 0
	\end{array}\right), \quad \epsilon_{\alpha\beta}=\left(\begin{array}{cc}
	0 & -1 \\
	1 & 0
\end{array}\right).
\end{eqnarray}
The term in the right-hand side of eq.~(\ref{eq:egcorres}) is the equation of motion of $\psi$ that vanishes when mapping to on-shell amplitudes, which indicates that mapping operators to on-shell amplitudes automatically removes operators involving the EOM and the CDC.

To obtain the Green's basis, we need to keep operators involving the EOM. Here we propose an off-shell amplitude formalism such that operators carrying all possible redundancies correspond to off-shell amplitudes by a one-to-one mapping. The one-to-one mapping from operators to off-shell amplitudes is introduced as,
\eq{\label{eq:offshellmap}
	\begin{array}{lcl}
		F_{{\tiny\rm L/R}\,i}	&	\rightarrow	&	\lambda_{i,0}\lambda_{i,0}/\tilde\lambda_{i,0}\tilde\lambda_{i,0}			\\
		\psi_i/\psi^{\dagger}_i	&	\rightarrow	&	\lambda_{i,0}/\tilde\lambda_{i,0}	\\
		\phi_i &	\rightarrow	& 1 \\
		D_{i,d_i}	&	\rightarrow	&	-i\lambda_{i,d_i}\tilde\lambda_{i,d_i}
	\end{array},
}
where $i$ in the subscript of each field labels the $i$th field in an operator, while $i$ in the subscript of a covariant derivative denotes that the covariant derivative acts on the $i$th field. $0$ in the subscript of a spinor indicates the spinor index corresponds to a field. $d_i$ indicates the order of covariant derivatives acting on the $i$th field, $d_i \in \{1,\dots,\hat{d}_i\}$, $\hat{d}_i \in \mathbb{Z}$, where $d_i=1$ labels the first covariant derivative acting on the $i$th field, and $d_i=2$ labels the second covariant derivative acting on that field, etc. In the following context, we will also use $x_i$ to label a spinor, $x_i \in \{0,1,\dots,\hat{d}_i\}$. $x_i=0$ indicates that the spinor corresponds to a field, while $x_i \neq 0$ indicates that the spinor corresponds to a covariant derivative, and the order of the derivative, among all derivatives acting on one specific field, is encoded in the index as explained above. The procedure to recover from the off-shell amplitudes to on-shell amplitudes is manifest by substituting all $\lambda_{i,x_i}$s of the $i$th building block with $\lambda_{i,0}=\lambda_i$ for all $i \in \{1,2,\dots,N\}$ in an off-shell amplitude makes the amplitude on-shell.

Here is an example to see how operators involving the EOM are kept in this off-shell amplitude formalism. For example,
\begin{equation}
    \psi_1^{\alpha} (\sigma^{\nu})_{\alpha\dot{\alpha}} D^{\mu} D_{\mu} D_{\nu} \psi_2^{\dagger\dot{\alpha}} = \dfrac{1}{2}\psi_1^{\alpha} D^{\beta\dot{\beta}} D_{\beta\dot{\beta}} D_{\alpha\dot{\alpha}} \psi_2^{\dagger\dot{\alpha}} \rightarrow -\dfrac{i}{2}\vev{1_02_1} \vev{2_32_2} [2_12_0] [2_32_2],
\end{equation}
where $\vev{i_{x_i}j_{x_j}}=\lambda^{\alpha}_{i,x_i}\lambda_{j,x_j\alpha}$ and $[i_{x_i}j_{x_j}]=\tilde\lambda_{i,x_i\dot{\alpha}}\tilde\lambda_{j,x_j}^{\dot{\alpha}}$. This amplitude does not vanish since $\lambda_{i,x_i}/\tilde\lambda_{i,x_i}$s with different subscripts $x_i$s are regarded as different spinors. It can be seen that the operator is kept while mapping to the amplitude, and it is straightforward to trace back and find the operator that is the reverse of the amplitude since the mapping is one-to-one.

The fields in the SMEFT are representations of the gauge group $SU(3)_C$, $SU(2)_W$ and $U(1)_Y$ as well.  The field contents of the SMEFT are listed in table~\ref{tab:SMEFT-field-content}. Since the representations of all fields in the SMEFT include trivial, fundamental, anti-fundamental and adjoint representation of the $SU(3)_C$ and the $SU(2)_W$ groups, the invariant tensors that contract with fields to form a singlet of the gauge groups must consist of the following basic invariant tensors:
\bea\label{eq:SU3tensors}
&& SU(3) : \quad f^{ABC},d^{ABC},\delta^{AB},(\lambda^A)_a^b,\epsilon_{abc},\epsilon^{abc}, \delta^a_b\\
&& SU(2) : \quad \epsilon^{IJK},\delta^{IJ},(\tau^I)_i^j,\epsilon_{ij},\epsilon^{ij},\delta^i_j,\label{eq:SU2tensors}
\eea
where $\{i,j,k,\cdots\}$ and $\{I,J,K,\cdots\}$ denote the indices of the (anti)fundamental and adjoint representation of the $SU(2)_W$ group, and $\{a,b,c,\cdots\}$ and $\{A,B,C,\cdots\}$ denote the indices of the (anti)fundamental and adjoint representation of the $SU(3)_C$ group. $\tau$ and $\lambda$ are the Pauli matrices and the Gell-Mann matrices.

\begin{table}[t]
	\begin{center}
		\begin{tabular}{|c|cc|ccc|ccc|}
			\hline
			\text{Fields} & $SU(2)_{l}\times SU(2)_{r}$	& $h$ & $SU(3)_{C}$ & $SU(2)_{W}$ & $U(1)_{Y}$ &  Flavor & $B$ & $L$ \tabularnewline
			\hline
			$G_{\rm L\alpha\beta}^A$   & $\left(1,0\right)$  & $-1$    & $\boldsymbol{8}$ & $\boldsymbol{1}$ & 0  & $1$ & 0 & 0 \tabularnewline
			$W_{\rm L\alpha\beta}^I$   & $\left(1,0\right)$  & $-1$           & $\boldsymbol{1}$ & $\boldsymbol{3}$ & 0  & $1$ & 0 & 0 \tabularnewline
			$B_{\rm L\alpha\beta}$   & $\left(1,0\right)$    & $-1$        & $\boldsymbol{1}$ & $\boldsymbol{1}$ & 0  & $1$ & 0 & 0 \tabularnewline
			\hline
			$L_{\alpha i}$     & $\left(\frac{1}{2},0\right)$  & $-\frac12$  & $\boldsymbol{1}$ & $\boldsymbol{2}$ & $-\frac12$  & $n_f$ & 0 & $1$ \tabularnewline
			$e_{_\mathbb{C}\alpha}$ & $\left(\frac{1}{2},0\right)$ & $-\frac12$   & $\boldsymbol{1}$ & $\boldsymbol{1}$ & $1$  & $n_f$ & 0 & $-1$ \tabularnewline
			$Q_{\alpha ai}$     & $\left(\frac{1}{2},0\right)$ & $-\frac12$   & $\boldsymbol{3}$ & $\boldsymbol{2}$ & $\frac16$  & $n_f$ & $\frac13$ & 0 \tabularnewline
			$u_{_\mathbb{C}\alpha}^a$ & $\left(\frac{1}{2},0\right)$ & $-\frac12$   & $\overline{\boldsymbol{3}}$ & $\boldsymbol{1}$ & $-\frac23$  & $n_f$ & $-\frac13$ & 0 \tabularnewline
			$d_{_\mathbb{C}\alpha}^a$ & $\left(\frac{1}{2},0\right)$ & $-\frac12$   & $\overline{\boldsymbol{3}}$ & $\boldsymbol{1}$ & $\frac13$  & $n_f$ & $-\frac13$ & $0$ \tabularnewline
			\hline
			$H_i$     & $\left(0,0\right)$&  0     & $\boldsymbol{1}$ & $\boldsymbol{2}$ & $\frac12$  & $1$ & 0 & 0 \tabularnewline
			\hline
		\end{tabular}
		\caption{\label{tab:SMEFT-field-content}
			The field contents of the standard model, along with their representations under the Lorentz and gauge groups.
			The number of flavors of each fermionic field is denoted by $n_f$, which is 3 in the SM. The baryon number $B$ and the lepton number $L$ of each field are also listed. All of the fields are accompanied with their Hermitian conjugates which are omitted, $(F_{\rm L \alpha\beta})^\dagger = F_{\rm R \dot\alpha\dot\beta}$ for gauge bosons, $(\psi_\alpha)^\dagger = (\psi^\dagger)_{\dot\alpha}$ for fermions, and $H^{\dagger}$ for the Higgs. }
	\end{center}
\end{table}

Before going further into the construction of the basis, we introduce some terminologies that are useful in the following. 
\begin{itemize}
	\item Class: The Lorentz irreducible representations denoted by a set of abstract fields $\{F_{\rm L}, \psi, \phi, \psi^{\dagger}, F_{\rm R}\}$ and covariant derivatives that can be Lorentz invariant form a Lorentz class.
	\item Type: Substituting fields of the SMEFT into the abstract fields of each Lorentz class that can be gauge invariant is a type. If the Hermitian conjugate of a type is the type itself, the type is called a real type, otherwise it is called a complex type.
	\item Term: For each type, the Lorentz and gauge invariant tensors with $m$ repeated fields in the type transform under certain representation of the symmetric group $S_m$. The decomposition of the representations into irreducible representations of the $S_m$ group gives the terms in that type.
	\item Operator: The irreducible representations of the $S_m$ are also irreducible representations of $SU(n_f)$ due to the Schur-Weyl duality, where $n_f$ denotes the flavor number of the repeated fields. Specifying each flavor index with $\{1,\cdots,n_f\}$ gives the independent flavor-specified operators in a term.
\end{itemize}
Furthermore, we also introduce several concepts for enumerating the operator basis~\cite{Li:2022tec}:
\bit
\item the y-basis: the complete and independent basis for Lorentz and gauge structures obtained by the young tableau method. 
\item the m-basis: the independent monomial basis with notations that are familiar to the phenomenology community. 
\item the p-basis: the re-organized basis in terms of irreducible representation of the permutation group for the flavor indices of 
the repeated fields. 
\item the f-basis: the final operator basis that are allowed by the number of flavors.
\eit

\section{Off-shell Construction of the Green's Basis}\label{sec:Green}

In this section, we introduce the procedure to obtain the Green's basis in the SMEFT at any mass dimension, which can be applied to general EFTs. In subsection~\ref{sec:class}, we summarize the method to enumerate all Lorentz classes at a given dimension. In subsection~\ref{sec:lorybasis} and \ref{sec:lormbasis}, we propose the off-shell construction of the Lorentz y-basis and m-basis with detailed examples. Then in subsection~\ref{sec:gauymbasis}, we review the method to obtain the gauge y-basis and m-basis. At last, we combine the off-shell Lorentz basis with the gauge basis to generate the off-shell p-basis and the off-shell f-basis where operators are presented as irreducible representations of the $SU(n_f)$ group of flavor indices in subsection~\ref{sec:pbasis}.


\subsection{Lorentz Classes in Green's Basis}\label{sec:class}

In this subsection, we briefly introduce the method of enumerating all classes at a given dimension $\mc{D}$, which has been elaborated in Ref.~\cite{Li:2020gnx}. For a given class, we introduce a state $\{h_1,h_2,\cdots,h_N\}$ and the number of derivatives in the class $n_D$, where $N$ denotes the number of fields in the class and $h_i$ denotes the helicity of the $i$th field, $i=1,\cdots,N$. The helicities $\{h_i\}$ are sorted in the ascending order. $n$ and $\tilde{n}$ are defined as half the total number of the left-handed spinors and the right-handed spinors in the class respectively,
\begin{equation}\label{eq:ntilden}
	n = \sum_{i,h_i<0}|h_i| + \frac{1}{2} n_D, \quad \tilde{n} = \sum_{i,h_i>0}|h_i| + \frac{1}{2} n_D.
\end{equation}
A general class can be present as $F_{\rm L}^{n_{-1}}\psi^{n_{-1/2}}\phi^{n_0}\psi^{{}_\dagger n_{1/2}}F_{\rm R}^{n_1}D^{n_D}$, and we need to consider all possible values of the tuple $\{n_{-1},n_{-1/2},n_0,n_{1/2},n_1,n_D\}$ with the following constraints:
\eq{\label{eq:lor-inv_constraint}
	& \tilde{n}+n = \mc{D}-N, \qquad \tilde{n}-n = \sum_ih_i \equiv h, \qquad \sum_{i=1}^N n_i = N,\\
	& 2n_{-1}+n_{-1/2} = \sum_i|h_i|-h = 2n-n_D, \quad 2n_{1}+n_{1/2} = \sum_i|h_i|+h = 2\tilde{n}-n_D, \\
	& \min(2n, 2\tilde{n}) \geq n_D \geq \max\begin{pmatrix} h-\sum_i|h_i|,\ \mod 2 \\ 4|\min h_i| - \sum_{h_i<0}2|h_i| \\ 4|\max h_i| - \sum_{h_i>0}2|h_i| \end{pmatrix}.
}

For the Green's basis, it should be noted that cases where $N\leq3$ and $n_D>0$ need to be taken into account, which automatically vanish if the on-shell construction is utilized while can be kept using the off-shell amplitude formalism.
All the classes in the dimension 8 Green's basis in the SMEFT are listed in table~\ref{tab:classes8}. 

\begin{table}[h]
	\eq{
		\begin{array}{ll|llll}
			\hline
			N		&	(n,\tilde{n})
			& 	& \text{Classes}	&	&	\\
			\hline
			2		&	(4,2)	& F_{\rm L}^2 D^4+\hc	&	&	&	\\
			&	(3,3)	&	F_{\rm L} F_{\rm R} D^4	&	\phi^2 D^6	&	\psi \psi^{\dagger} D^5	\\
			\hline
			3		&	(4,1)	& F_{\rm L}^3 D^2 +\hc	&	&	&	\\
			&	(3,2)	&	F_{\rm L}^2 F_{\rm R} D^2 +\hc	&	F_{\rm L} \phi^2 D^4+\hc		&	\psi^2\phi D^4+\hc	& F_{\rm L} \psi \psi^{\dagger} D^3 +\hc	\\
			\hline
			4		&	(4,0)	&	F_{\rm L}^4+\hc	&	&	&	\\
			&	(3,1)	&	F_{\rm L}^2\psi\psi^{\dagger}D+\hc		&	\psi^4D^2+\hc				&	F_{\rm L}\psi^2\phi D^2+\hc	&	F_{\rm L}^2\phi^2 D^2+\hc	\\
			&	(2,2)	&	F_{\rm L}^2F_{\rm R}^2	&	F_{\rm L}F_{\rm R}\psi\psi^{\dagger}D	&	\psi^2\psi^{\dagger 2}D^2	&	F_{\rm R}\psi^2\phi D^2+\hc	\\
			&			&	F_{\rm L}F_{\rm R}\phi^2 D^2	&	\psi\psi^{\dagger}\phi^2D^3	&	\phi^4D^4 	& \\
			\hline
			5		&	(3,0)	&	F_{\rm L}\psi^4+\hc	&	F_{\rm L}^2\psi^2\phi+\hc	&	F_{\rm L}^3\phi^2+\hc	&	\\
			&	(2,1)	&	F_{\rm L}\psi^2\psi^{\dagger 2}+\hc	&	F_{\rm L}^2\psi^{\dagger2}\phi+\hc	&	\psi^3\psi^\dagger\phi D+\hc	&	F_{\rm L}\psi\psi^\dagger\phi^2 D+\hc \\
			&			&	\psi^2\phi^3D^2+\hc	&	F_{\rm L}\phi^4D^2+\hc	&	&	\\
			\hline
			6		&	(2,0)	&	\psi^4\phi^2+\hc	&	F_{\rm L}\psi^2\phi^3+\hc	&	F_{\rm L}^2\phi^4+\hc	&	\\
			&	(1,1)	&	\psi^2\psi^{\dagger 2}\phi^2	&	\psi\psi^\dagger\phi^4D	&	\phi^6D^2	& \\
			\hline
			7		&	(1,0)	&	\psi^2\phi^5+\hc	&	&	&	\\
			\hline
			8		&	(0,0)	&	\phi^8	&	&	&	\\
			\hline
		\end{array}\notag
	}
	\caption{All the classes at dimension 8 in the Green's basis in the SMEFT. Classes with $N=2,3$ exist in the Green's basis while vanish in the on-shell basis of dimension-8 operators in the SMEFT. We only list classes with $n\geq\tilde{n}$, since all the classes with $n<\tilde{n}$ are Hermitian conjugate of certain classes listed here (denoted as $+\hc$).}\label{tab:classes8}
\end{table}

\subsection{Off-shell Lorentz Y-basis}\label{sec:lorybasis}

In this subsection, we will introduce the method to obtain the complete and independent basis of the Lorentz structures in a certain class. First of all, by distributing the derivatives to each field in the class and find all possible sets of spinors in the off-shell amplitude formalism, we generate all possible contractions by pairing up the spinors for each set of spinors. Traversing all possible sets of spinors and generating all contractions of spinors will give us an over-complete basis of the off-shell amplitudes in the class. At last we impose the redundancy relations among the off-shell amplitudes, which come from the redundancy relations among operators, and remove the redundant ones. Since we want to keep operators related by the field re-definition, operators differed by the EOM should be regarded as independent. Therefore the redundancies include the IBP\footnote{In fact, the IBP redundancy can be removed in the stage of distributing the derivatives to the fields, as we will elaborate later in the example.}, 
the CDC and some algebraic identities, which are the Schouten identity and the Bianchi identity in the off-shell amplitude formalism. 

Here we summarize the redundancies and our method to remove these redundancies.
\begin{itemize}
	\item The IBP relation among operators can be written explicitly in the off-shell amplitude formalism as
	\begin{equation}\label{eq:IBP}
		|i_{\hat{d}_i}\rangle[i_{\hat{d}_i}|=-\sum^N_{j=1,j\neq i} |j_{\hat{d}_j+1}\rangle[j_{\hat{d}_j+1}|,
	\end{equation}
    where $\hat{d}_i \equiv \max d_i$. Eq.~(\ref{eq:IBP}) shows that the outermost derivative acting on the $i$th field can be moved to the outermost of the other $N-1$ fields. In the on-shell construction~\cite{Li:2020xlh}, there are several reduction rules to remove the IBP since the EOM should be taken into account when dealing with the IBP,
    \eq{\label{eq:IBPold1}
		\langle i1\rangle[1j]=-\sum_{k=2}^{N}\langle ik\rangle[kj],
	}
	\eq{\label{eq:IBPold2}
		[1|p_2|i\rangle=-\sum_{k=3}^N[1|p_k|i\rangle,\quad &\langle 1|p_2|i]=-\sum_{k=3}^N\langle 1|p_k|i] ,\\
		[1|p_3|2\rangle=-\sum_{k=4}^N[1|p_k|2\rangle,\quad &\langle 1|p_3|2]=-\sum_{k=4}^N\langle 1|p_k|2] ,
	}
	\eq{\label{eq:IBPold3}
		p_1^2 = 2\sum_{i,j\neq1} p_i\cdot p_j = 0 .
	}
	Eq.~(\ref{eq:IBPold2}) and eq.~(\ref{eq:IBPold3}) are supplementaries of eq.~(\ref{eq:IBPold1}), as $\vev{11}$ and $[11]$, which correspond to the EOM and automatically vanish, could appear in eq.~(\ref{eq:IBPold1}) and substituting the left-hand side to the right-hand side of eq.~(\ref{eq:IBPold1}) will not be enough to remove the IBP completely.
    However, in the off-shell construction, we only need to modify eq.~(\ref{eq:IBPold1}) to eq.~(\ref{eq:IBP}) in the off-shell amplitude formalism and keep the terms on the right-hand side of eq.~(\ref{eq:IBP}) to remove the IBP of $D_{i,\hat{d}_i}$ as long as the term on the left-hand side of eq.~(\ref{eq:IBP}) does not vanish due to the EOM. In fact, all derivatives acting on a certain field can be moved to other fields one by one using the corresponding IBP relations of these derivatives, which suggests that choosing the amplitudes where no derivative acts on that field as the independent ones from all possible amplitudes completely removes the IBP redundancy. In this work, we always choose that no derivative acts on the first field in an operator where the fields are sorted by their helicities in the ascending order.
    
    \item The Schouten identity of the left-handed spinors can be presented in the off-shell amplitude formalism as
    \begin{equation}\label{eq:Schouten}
    	\langle i_{x_i} l_{x_l} \rangle \langle j_{x_j} k_{x_k} \rangle = - \langle i_{x_i} j_{x_j} \rangle \langle k_{x_k} l_{x_l} \rangle + \langle i_{x_i} k_{x_k} \rangle \langle j_{x_j} l_{x_l} \rangle,
    \end{equation}
    and the Schouten identity of the right-handed spinors is the same except that the angular brackets are replaced by square brackets. In Ref.~\cite{Li:2020xlh}, the Schouten identity for any four different (left-handed) on-shell spinors in the on-shell amplitudes is given as
    \begin{equation}\label{eq:Schoutenonshell}
    	\langle i l \rangle \langle j k \rangle = - \langle i j \rangle \langle k l \rangle + \langle i k \rangle \langle j l \rangle, \quad i < j < k < l.
    \end{equation}
    Unlike eq.~(\ref{eq:Schoutenonshell}),
    the Schouten identity eq.~(\ref{eq:Schouten}) holds among any four different left(right)-handed off-shell spinors, including the cases where two or more of the $i,j,k,l$ are the same while the corresponding $x_i,x_j,x_k,x_l$ are not. To deal with those cases, here we set up the order of labels of off-shell spinors as
    \begin{eqnarray}\label{eq:spindorder}
    	i_{x_i} < j_{x_j}, \quad &\text{if $i<j$, or $i=j$ and $x_i<x_j$},
    \end{eqnarray}
    and we always replace the left side of eq.~(\ref{eq:Schouten}) by the right side of the equation when $i_{x_i} < j_{x_j} < k_{x_k} < l_{x_l}$.
    
    \item The CDC relation is $[D_{\mu}, D_{\nu}]=-iF_{\mu\nu}$, which implies that all derivatives acting on one field are totally symmetric since any two antisymmetric ones can be replaced by a field strength tensor. In the off-shell amplitude formalism, that means the amplitude should be totally symmetric under any permutation of the set of spinor indices $\{i_{d_i}|d_i \in \{1,2,\cdots,\hat{d_i}\}\}$ for all $i$. We utilize the Young symmetrizer $\mc{Y}\left[{\scriptsize \young({{i_1}}{{i_2}})\cdots\young({{i_{\hat{d}_i}}})}\right]$ to symmetrize the derivatives acting on the $i$th field for all fields in an operator.
    
    \item The Bianchi identity of a gauge field $F_{\mu\nu}$ reads
    \begin{eqnarray}\label{eq:Bianchi1}
    	\epsilon^{\mu \nu \rho \sigma} D_{\nu} F_{\rho \sigma}=0,
    \end{eqnarray}
    which implies
    \eq{\label{eq:Bianchi2}
    	D_{\mu} F^{\mu\nu}_{\rm L} -D_{\mu} F^{\mu\nu}_{\rm R}=-iD_{\mu} \tilde{F}^{\mu\nu} =0,
    }
    or written in spinor indices as
    \eq{\label{eq:Bianchi3}
    	D_{\beta\dot{\alpha}} F_{\rm L}{}_{\alpha}{}^{\beta} - D_{\alpha\dot{\beta}} F_{\rm R}{}^{\dot{\beta}}{ }_{\dot{\alpha}} &= \frac{1}{2} D_{\beta\dot{\beta}} \tilde{F}_{\mu\nu} (\sigma^{\mu \nu})_{\alpha}{}^{\beta} \delta^{\dot{\beta}}{ }_{\dot{\alpha}} - \frac{1}{2} D_{\beta\dot{\beta}} \tilde{F}_{\mu\nu} (\bar{\sigma}^{\mu \nu})^{\dot{\beta}}{ }_{\dot{\alpha}} \delta_{\alpha}{ }^{\beta} \\
    	&= -i D_{\rho} \tilde{F}_{\mu\nu} \sigma^{\rho}_{\beta\dot{\beta}} \sigma^{\mu\beta\dot{\beta}} \sigma^{\nu}_{\alpha\dot{\alpha}} \\
    	&= -2iD^{\mu} \tilde{F}_{\mu\nu} \sigma^{\nu}_{\alpha\dot{\alpha}} \\
    	&=0,
    }
    where $\tilde{F}^{\mu \nu}=\frac{1}{2} \epsilon^{\mu \nu \rho \eta} F_{\rho \eta}$.
    Eq.~(\ref{eq:Bianchi1}-\ref{eq:Bianchi3}) indicate that the EOM of $F_{\rm L}$ and $F_{\rm R}$ are not independent because the EOM of $\tilde{F}$ vanishes due to the Bianchi identity. In this case, there is an alternative to choose whether to keep the EOM of $F_{\rm L}$ or the EOM of $F_{\rm R}$, and we choose to keep the EOM of $F_{\rm L}$, which means $F_{\rm R}$ is taken to be on-shell. Since $F_{\rm R}$ will not be the first field in a class in our notation unless the class only involves $F_{\rm R}$, in which case all fields are on-shell and the independent Lorentz structures of the class can be obtained through the on-shell construction, the previous method of removing the IBP still holds.  
\end{itemize}

We take the class $F_{\rm L 1} \psi_2 \psi_3 \phi_4 D^2$ as an example to illustrate the method. From the mapping eq.~(\ref{eq:offshellmap}), the spinors in this class must include $\lambda_{1,0}^2\lambda_{2,0}\lambda_{3,0}$. Since there are $C_4^1+C_4^2=10$ ways to arrange the two derivatives acting on the four fields, all possible sets of spinors that come from the derivatives in the class are
\eq{\label{eq:egspinor}
    \lambda_{1,1} \lambda_{1,2} \tilde\lambda_{1,1} \tilde\lambda_{1,2}, \quad \lambda_{2,1} \lambda_{2,2} \tilde\lambda_{2,1} \tilde\lambda_{2,2}, \quad
    \lambda_{3,1} \lambda_{3,2} \tilde\lambda_{3,1} \tilde\lambda_{3,2}, \quad \lambda_{4,1} \lambda_{4,2} \tilde\lambda_{4,1} \tilde\lambda_{4,2}, \quad
    \lambda_{1,1} \lambda_{2,1} \tilde\lambda_{1,1} \tilde\lambda_{2,1}, \\
    \lambda_{1,1} \lambda_{3,1} \tilde\lambda_{1,1} \tilde\lambda_{3,1}, \quad \lambda_{1,1} \lambda_{4,1} \tilde\lambda_{1,1} \tilde\lambda_{4,1}, \quad \lambda_{2,1} \lambda_{3,1} \tilde\lambda_{2,1} \tilde\lambda_{3,1}, \quad \lambda_{2,1} \lambda_{4,1} \tilde\lambda_{2,1} \tilde\lambda_{4,1}, \quad \lambda_{3,1} \lambda_{4,1} \tilde\lambda_{3,1} \tilde\lambda_{4,1}.
}
Before considering how those spinors contract, let us pay some attention to the IBP redundancy.
From the discussion below eq.~(\ref{eq:IBP}), we can choose that the independent amplitudes are those where no derivative acts on the first field, and then only the following 6 sets of the 10 sets of spinors in (\ref{eq:egspinor}) should be kept
\eq{\label{eq:egspinorIBP}
     \lambda_{2,1} \lambda_{2,2} \tilde\lambda_{2,1} \tilde\lambda_{2,2}, \quad
    \lambda_{3,1} \lambda_{3,2} \tilde\lambda_{3,1} \tilde\lambda_{3,2}, \quad \lambda_{4,1} \lambda_{4,2} \tilde\lambda_{4,1} \tilde\lambda_{4,2}, \\
    \lambda_{2,1} \lambda_{3,1} \tilde\lambda_{2,1} \tilde\lambda_{3,1}, \quad \lambda_{2,1} \lambda_{4,1} \tilde\lambda_{2,1} \tilde\lambda_{4,1}, \quad \lambda_{3,1} \lambda_{4,1} \tilde\lambda_{3,1} \tilde\lambda_{4,1}.
}

Now we are ready to move on to the contractions of the spinors. Taking account of the spinors $\lambda_{1,0}^2\lambda_{2,0}\lambda_{3,0}$ from the fields in the class $F_{\rm L 1} \psi_2 \psi_3 \phi_4 D^2$, there are 6 different contractions for each set of spinors in (\ref{eq:egspinorIBP}).
For example, the following 6 contractions of $\lambda_{1,0}^2\lambda_{2,0}\lambda_{2,1} \lambda_{2,2} \lambda_{3,0} \tilde\lambda_{2,1} \tilde\lambda_{2,2}$  are different considering the antisymmetric nature of the brackets $\vev{i_{x_i}j_{x_j}}=-\vev{j_{x_j}i_{x_i}}$ and $[i_{x_i}j_{x_j}]=-[j_{x_j}i_{x_i}]$,
\eq{\label{eq:egspcontract}
\vev{1_0 2_0} \vev{1_0 2_1} \vev{2_2 3_0} [2_1 2_2], \quad \vev{1_0 2_0} \vev{1_0 2_2} \vev{2_1 3_0} [2_1 2_2], \quad \vev{1_0 2_0} \vev{1_0 3_0} \vev{2_1 2_2} [2_1 2_2], \\
\vev{1_0 2_1} \vev{1_0 2_2} \vev{2_0 3_0} [2_1 2_2], \quad \vev{1_0 2_1} \vev{1_0 3_0} \vev{2_0 2_2} [2_1 2_2], \quad \vev{1_0 2_2} \vev{1_0 3_0} \vev{2_0 2_1} [2_1 2_2].
}
However, the 6 amplitudes in (\ref{eq:egspcontract}) are not independent due to the Schouten identity.
In the convention we adopted below eq.~(\ref{eq:Schouten}), the 6 amplitudes in (\ref{eq:egspcontract}) are left with 3 of them after removing the redundancy of the Schouten identity,
\begin{eqnarray}\label{eq:egspschouten}
    \vev{1_0 2_0} \vev{1_0 2_1} \vev{2_2 3_0} [2_1 2_2], \quad \vev{1_0 2_0} \vev{1_0 2_2} \vev{2_1 3_0} [2_1 2_2], \quad \vev{1_0 2_1} \vev{1_0 2_2} \vev{2_0 3_0} [2_1 2_2].
\end{eqnarray}
Up to now we can conclude that the above 3 amplitudes span a algebraically independent and complete basis in the space of amplitudes formed by $\lambda_{1,0}^2\lambda_{2,0}\lambda_{2,1} \lambda_{2,2} \lambda_{3,0} \tilde\lambda_{2,1} \tilde\lambda_{2,2}$. This step of removing the Schouten identity can also be achieved using the Young tableaux method. Consider all possible ways to fill a $2\times 3$ Young diagram and a $2\times 1$ Young diagram with $\{1_0,1_0,2_0,2_1,2_2,3_0\}$ and $\{2_1,2_2\}$ respectively and get semi-standard Young tableaux,
\begin{eqnarray}\label{eq:ytab1}
    &\young({{1_0}}{{1_0}}{{2_2}},{{2_0}}{{2_1}}{{3_0}})  \quad \young({{1_0}}{{1_0}}{{2_1}},{{2_0}}{{2_2}}{{3_0}}) \quad \young({{1_0}}{{1_0}}{{2_0}},{{2_1}}{{2_2}}{{3_0}}) \ , \\
    &\young({{2_1}},{{2_2}}) \ .\label{eq:ytab2}
\end{eqnarray}
Here the semi-standard Young tableau means a Young tableau that the numbers in each row are non-descending and the numbers in each column are ascending in the order defined in (\ref{eq:spindorder}). It is straightforward to check that contracting the labels of spinors in each column of the Young tableaux in (\ref{eq:ytab1}) and (\ref{eq:ytab2}) with angular brackets and square brackets gives the same result in (\ref{eq:egspschouten}).

Although the 3 amplitudes in (\ref{eq:egspschouten}) are algebraically independent, the redundancy caused by the CDC is still there. For instance, we can find the following relations among the amplitudes through observation,
\begin{eqnarray}
    \vev{1_0 2_0} \vev{1_0 2_1} \vev{2_2 3_0} [2_1 2_2] \xLeftrightarrow{\text{exchange $2_1,2_2$}} -\vev{1_0 2_0} \vev{1_0 2_2} \vev{2_1 3_0} [2_1 2_2], \\
    \vev{1_0 2_1} \vev{1_0 2_2} \vev{2_0 3_0} [2_1 2_2] \xLeftrightarrow{\text{exchange $2_1,2_2$}} -\vev{1_0 2_1} \vev{1_0 2_2} \vev{2_0 3_0} [2_1 2_2].
\end{eqnarray}
To remove the redundancy of the CDC, we utilize the Young symmetrizer $\mc{Y}\left[{\scriptsize \young({{2_1}}{{2_2}})}\right]$ to symmetrize the spinor indices $2_1$ and $2_2$ that label the first derivative and the second derivative acting on the field $\psi_2$,
\eq{
    \mc{Y}\left[{\scriptsize \young({{2_1}}{{2_2}})}\right] \vev{1_0 2_0} \vev{1_0 2_1} \vev{2_2 3_0} [2_1 2_2] &= \frac{1}{2} (\vev{1_0 2_0} \vev{1_0 2_1} \vev{2_2 3_0} [2_1 2_2] + \vev{1_0 2_0} \vev{1_0 2_2} \vev{2_1 3_0} [2_2 2_1]), \\
    \mc{Y}\left[{\scriptsize \young({{2_1}}{{2_2}})}\right] \vev{1_0 2_0} \vev{1_0 2_2} \vev{2_1 3_0} [2_1 2_2] &= \frac{1}{2} (\vev{1_0 2_0} \vev{1_0 2_2} \vev{2_1 3_0} [2_1 2_2] + \vev{1_0 2_0} \vev{1_0 2_1} \vev{2_2 3_0} [2_2 2_1]), \\
    \mc{Y}\left[{\scriptsize \young({{2_1}}{{2_2}})}\right] \vev{1_0 2_1} \vev{1_0 2_2} \vev{2_0 3_0} [2_1 2_2] &= \frac{1}{2} (\vev{1_0 2_1} \vev{1_0 2_2} \vev{2_0 3_0} [2_1 2_2] + \vev{1_0 2_2} \vev{1_0 2_1} \vev{2_0 3_0} [2_2 2_1]),
}
where $\mc{Y}\left[{\scriptsize \young(ij)}\right]=\frac{1}{2}(E+(i \ j))$ is the totally symmetric Young symmetrizer of the $S_2$ group.
Recall that the 3 amplitudes in (\ref{eq:egspschouten}) span a algebraically independent and complete basis in the space of amplitudes formed by $\lambda_{1,0}^2\lambda_{2,0}\lambda_{2,1} \lambda_{2,2} \lambda_{3,0} \tilde\lambda_{2,1} \tilde\lambda_{2,2}$, which means any amplitude in the space can be reduced into the basis using the antisymmetric nature of brackets and the Schouten identity. Thus we obtain
\begin{eqnarray}\label{eq:egspCDC}
    \mc{Y}\left[{\scriptsize \young({{2_1}}{{2_2}})}\right] \left(\begin{array}{c}
         \vev{1_0 2_0} \vev{1_0 2_1} \vev{2_2 3_0} [2_1 2_2] \\
         \vev{1_0 2_0} \vev{1_0 2_2} \vev{2_1 3_0} [2_1 2_2] \\
         \vev{1_0 2_1} \vev{1_0 2_2} \vev{2_0 3_0} [2_1 2_2]
    \end{array}\right) = \left(\begin{array}{ccc}
        \Rd{\frac{1}{2}} & \Rd{-\frac{1}{2}} & \Rd0 \\
        -\frac{1}{2} & \frac{1}{2} & 0 \\
        0 & 0 & 0
    \end{array}\right)
    \left(\begin{array}{c}
         \vev{1_0 2_0} \vev{1_0 2_1} \vev{2_2 3_0} [2_1 2_2] \\
         \vev{1_0 2_0} \vev{1_0 2_2} \vev{2_1 3_0} [2_1 2_2] \\
         \vev{1_0 2_1} \vev{1_0 2_2} \vev{2_0 3_0} [2_1 2_2]
    \end{array}\right),
\end{eqnarray}
from which we can see that there is only one independent amplitude after considering the CDC since the matrix is rank-1, and the independent amplitudes can be chosen to be the independent rows of the matrix. Here we choose the red row of the matrix in eq.~(\ref{eq:egspCDC}) as the one independent amplitude of all amplitudes formed by $\lambda_{1,0}^2\lambda_{2,0}\lambda_{2,1} \lambda_{2,2} \lambda_{3,0} \tilde\lambda_{2,1} \tilde\lambda_{2,2}$, that is,
\begin{equation}
    \mc{Y}\left[{\scriptsize \young({{2_1}}{{2_2}})}\right] \vev{1_0 2_0} \vev{1_0 2_1} \vev{2_2 3_0} [2_1 2_2].
\end{equation}
Traversing all sets of spinors from the derivatives in eq.~(\ref{eq:egspinorIBP}) along with $\lambda_{1,0}^2\lambda_{2,0}\lambda_{3,0}$ from the fields, and following the procedure above, we obtain a basis of off-shell amplitudes in class $F_{\rm L 1} \psi_2 \psi_3 \phi_4 D^2$ as
\eq{\label{eq:ybasis1}
    \mc{B}^{\text{(off-shell) $y$}}_{F_{\rm L } \psi^2 \phi D^2,1} &= \mc{Y}\left[{\scriptsize \young({{2_1}}{{2_2}})}\right] \vev{1_0 2_0} \vev{1_0 2_1} \vev{2_2 3_0} [2_1 2_2], \\
    \mc{B}^{\text{(off-shell) $y$}}_{F_{\rm L} \psi^2 \phi D^2,2} &= \mc{Y}\left[{\scriptsize \young({{3_1}}{{3_2}})}\right] \vev{1_0 2_0} \vev{1_0 3_0} \vev{3_1 3_2} [3_1 3_2], \\
    \mc{B}^{\text{(off-shell) $y$}}_{F_{\rm L} \psi^2 \phi D^2,3} &= \mc{Y}\left[{\scriptsize \young({{4_1}}{{4_2}})}\right] \vev{1_0 2_0} \vev{1_0 3_0} \vev{4_1 4_2} [4_1 4_2], \\
    \mc{B}^{\text{(off-shell) $y$}}_{F_{\rm L} \psi^2 \phi D^2,4} &= \vev{1_0 2_0} \vev{1_0 2_1} \vev{3_0 3_1} [2_1 3_1], \\
    \mc{B}^{\text{(off-shell) $y$}}_{F_{\rm L} \psi^2 \phi D^2,5} &= \vev{1_0 2_0} \vev{1_0 3_0} \vev{2_1 3_1} [2_1 3_1], \\
    \mc{B}^{\text{(off-shell) $y$}}_{F_{\rm L} \psi^2 \phi D^2,6} &= \vev{1_0 2_1} \vev{1_0 3_0} \vev{2_0 3_1} [2_1 3_1], \\
    \mc{B}^{\text{(off-shell) $y$}}_{F_{\rm L} \psi^2 \phi D^2,7} &= \vev{1_0 2_0} \vev{1_0 2_1} \vev{3_0 4_1} [2_1 4_1], \\
    \mc{B}^{\text{(off-shell) $y$}}_{F_{\rm L} \psi^2 \phi D^2,8} &= \vev{1_0 2_0} \vev{1_0 3_0} \vev{2_1 4_1} [2_1 4_1], \\
    \mc{B}^{\text{(off-shell) $y$}}_{F_{\rm L} \psi^2 \phi D^2,9} &= \vev{1_0 2_1} \vev{1_0 3_0} \vev{2_0 4_1} [2_1 4_1], \\
    \mc{B}^{\text{(off-shell) $y$}}_{F_{\rm L} \psi^2 \phi D^2,10} &= \vev{1_0 2_0} \vev{1_0 3_0} \vev{3_1 4_1} [3_1 4_1], \\
    \mc{B}^{\text{(off-shell) $y$}}_{F_{\rm L} \psi^2 \phi D^2,11} &= \vev{1_0 2_0} \vev{1_0 3_1} \vev{3_0 4_1} [3_1 4_1], \\
    \mc{B}^{\text{(off-shell) $y$}}_{F_{\rm L} \psi^2 \phi D^2,12} &= \vev{1_0 3_0} \vev{1_0 3_1} \vev{2_0 4_1} [3_1 4_1].
}
The basis is called the off-shell y-basis, in analogy to the on-shell amplitude basis in Ref.~\cite{Li:2020xlh} called the y-basis. The off-shell amplitudes in eq.~(\ref{eq:ybasis1}) can be easily translated to operators by the one-to-one mapping (\ref{eq:offshellmap}), for instance,
\begin{eqnarray}\label{eq:egyoungCDC}
	\mc{Y}\left[{\scriptsize \young({{2_1}}{{2_2}})}\right] \vev{1_0 2_0} \vev{1_0 2_1} \vev{2_2 3_0} [2_1 2_2] \rightarrow \dfrac{1}{2} (F_{\rm L1}{}^{\alpha\beta} D^{\gamma\dot{\alpha}} D_{\beta\dot{\alpha}} \psi_{2\alpha} \psi_{3\gamma} \phi_4 + F_{\rm L1}{}^{\alpha\beta} D_{\beta\dot{\alpha}} D^{\gamma\dot{\alpha}} \psi_{2\alpha} \psi_{3\gamma} \phi_4),
\end{eqnarray}
We can recognize the above operator as either $F_{\rm L1}{}^{\alpha\beta} D^{\gamma\dot{\alpha}} D_{\beta\dot{\alpha}} \psi_{2\alpha} \psi_{3\gamma} \phi_4$ or $F_{\rm L1}{}^{\alpha\beta} D_{\beta\dot{\alpha}} D^{\gamma\dot{\alpha}} \psi_{2\alpha} \psi_{3\gamma} \phi_4$ since the three operators only differ by the CDC of the field $\psi_2$. In the following we will just omit the Young symmetrizers that symmetrize the derivatives, 
but one should always remember that all derivatives acting on a field are symmetrized in an operator basis.

The above procedure of constructing the off-shell amplitude basis of the class $F_{\rm L 1} \psi_2 \psi_3 \phi_4 D^2$ involves removing the IBP, the Schouten identity and the CDC redundancies, yet the Bianchi identity is not mentioned since the class $F_{\rm L 1} \psi_2 \psi_3 \phi_4 D^2$ does not involve $F_{\rm R}$. In the following we will give an example of the Hermitian conjugate of the above example $F_{\rm L 1} \psi_2 \psi_3 \phi_4 D^2$, that is, $ \phi_1 \psi^{\dagger}_2 \psi^{\dagger}_3 F_{\rm R 4} D^2$, where the fields are sorted by their helicities in the ascending order.

Similarly as the example $F_{\rm L 1} \psi_2 \psi_3 \phi_4 D^2$, the spinors of the fields in the class $ \phi_1 \psi^{\dagger}_2 \psi^{\dagger}_3 F_{\rm R 4} D^2$ are $\tilde{\lambda}_{2,0}\tilde{\lambda}_{3,0}\tilde{\lambda}^2_{4,0}$, and all possible sets of spinors of the derivatives modulo the IBP are
\eq{
	\lambda_{2,1} \lambda_{2,2} \tilde\lambda_{2,1} \tilde\lambda_{2,2}, \quad
	\lambda_{3,1} \lambda_{3,2} \tilde\lambda_{3,1} \tilde\lambda_{3,2}, \quad \lambda_{4,1} \lambda_{4,2} \tilde\lambda_{4,1} \tilde\lambda_{4,2}, \\
	\lambda_{2,1} \lambda_{3,1} \tilde\lambda_{2,1} \tilde\lambda_{3,1}, \quad \lambda_{2,1} \lambda_{4,1} \tilde\lambda_{2,1} \tilde\lambda_{4,1}, \quad \lambda_{3,1} \lambda_{4,1} \tilde\lambda_{3,1} \tilde\lambda_{4,1}.
}
To take a field $F_{\rm R}{}_i$ on shell, all $\tilde\lambda_{i,0}$s of the field $F_{\rm R}{}_i$ and all $\lambda_{i,d_i}/\tilde\lambda_{i,d_i}$s of the derivatives acting on the field should be substituted to $\lambda_{i,0}/\tilde\lambda_{i,0}$, which is equivalent to $\lambda_i/\tilde\lambda_i$ in the on-shell amplitude formalism, as mentioned in section~\ref{sec:offshell}. Then the procedure elaborated in the former example can be applied to find all independent amplitudes.

In this example, we simply apply the replacements $\lambda_{4,d_4} \rightarrow \lambda_{4,0}$ and $\tilde{\lambda}_{4,d_4} \rightarrow \tilde{\lambda}_{4,0}$, and the above sets of spinors become
\eq{\label{eq:eg2spinorIBP}
	\lambda_{2,1} \lambda_{2,2} \tilde\lambda_{2,1} \tilde\lambda_{2,2}, \quad
	\lambda_{3,1} \lambda_{3,2} \tilde\lambda_{3,1} \tilde\lambda_{3,2}, \quad \lambda_{4,0} \lambda_{4,0} \tilde\lambda_{4,0} \tilde\lambda_{4,0}, \\
	\lambda_{2,1} \lambda_{3,1} \tilde\lambda_{2,1} \tilde\lambda_{3,1}, \quad \lambda_{2,1} \lambda_{4,0} \tilde\lambda_{2,1} \tilde\lambda_{4,0}, \quad \lambda_{3,1} \lambda_{4,0} \tilde\lambda_{3,1} \tilde\lambda_{4,0}.
} 
If one consider the set of spinors $\lambda_{4,0} \lambda_{4,0} \tilde\lambda_{4,0} \tilde\lambda_{4,0}$ along with $\tilde{\lambda}_{2,0}\tilde{\lambda}_{3,0}\tilde{\lambda}^2_{4,0}$, there is no non-vanishing contraction of these spinors since $\vev{4_04_0}=0$, which suggests that all operators where the two derivatives acting on the field $F_{\rm R 4}$ involve the EOM of $F_{\rm R 4}$, thus are redundant due to the Bianchi identity. A non-vanishing example is the set of spinors $\lambda_{2,1} \lambda_{4,0}\tilde{\lambda}_{2,0}\tilde\lambda_{2,1}\tilde{\lambda}_{3,0}\tilde{\lambda}^3_{4,0}$. The only independent amplitude from contractions of the spinors is
\eq{
	\vev{2_14_0} [2_04_0][2_14_0][3_04_0],
}
which can be deduced from the Young tableaux method or simply listing all possible contractions and removing redundancy of the Schouten identity.
Applying the above method to all sets of spinors in (\ref{eq:eg2spinorIBP}) will give the off-shell y-basis of $ \phi_1 \psi^{\dagger}_2 \psi^{\dagger}_3 F_{\rm R 4} D^2$,
\eq{\label{eq:ybasis2}
	\mc{B}^{\text{(off-shell) $y$}}_{\phi \psi^{\dagger 2} F_{\rm R} D^2,1} &= \vev{2_1 2_2} [2_0 2_1] [2_2 4_0] [3_0 4_0], \\
	\mc{B}^{\text{(off-shell) $y$}}_{\phi \psi^{\dagger 2} F_{\rm R} D^2,2} &= \vev{3_1 3_2} [2_0 3_1] [3_0 4_0] [3_2 4_0], \\
	\mc{B}^{\text{(off-shell) $y$}}_{\phi \psi^{\dagger 2} F_{\rm R} D^2,3} &= \vev{2_1 3_1} [2_0 2_1] [3_0 4_0] [3_1 4_0], \\
	\mc{B}^{\text{(off-shell) $y$}}_{\phi \psi^{\dagger 2} F_{\rm R} D^2,4} &= \vev{2_1 3_1} [2_0 3_0] [2_1 4_0] [3_1 4_0], \\
	\mc{B}^{\text{(off-shell) $y$}}_{\phi \psi^{\dagger 2} F_{\rm R} D^2,5} &= \vev{2_1 3_1} [2_0 3_1] [2_1 4_0] [3_0 4_0], \\
	\mc{B}^{\text{(off-shell) $y$}}_{\phi \psi^{\dagger 2} F_{\rm R} D^2,6} &= \vev{2_14_0} [2_04_0][2_14_0][3_04_0], \\
	\mc{B}^{\text{(off-shell) $y$}}_{\phi \psi^{\dagger 2} F_{\rm R} D^2,7} &= \vev{3_14_0} [2_04_0][3_04_0][3_14_0],
}
where Young symmetrizers for spinor indices of the derivatives acting on each field are omitted, as we discussed below eq.~(\ref{eq:egyoungCDC}). The number of independent amplitudes in the off-shell y-basis of $ \phi_1 \psi^{\dagger}_2 \psi^{\dagger}_3 F_{\rm R 4} D^2$, eq.(\ref{eq:ybasis2}), is obviously less than that in the off-shell y-basis of the Hermitian conjugate of $F_{\rm L 1} \psi_2 \psi_3 \phi_4 D^2$,  eq.(\ref{eq:ybasis1}), because operators involve the EOM of $F_{\rm R}$ are dropped while operators involve the EOM of $F_{\rm L}$ are kept in the convention we chose.

If one want to use the mapping (\ref{eq:offshellmap}) to find the corresponding operator basis of the off-shell amplitude basis eq.(\ref{eq:ybasis2}), one need to retrieve the spinor indices of the derivatives for the field $F_{\rm R}$. The choice to retrieve the spinor indices of the derivatives can be arbitrary since different retrievals only differ by the EOM and the CDC of the field $F_{\rm R}$. For example, after retrieving the spinors of the derivatives in $\vev{2_14_0} [2_04_0][2_14_0][3_04_0]$, the amplitude can be the following 3 alternatives
\begin{eqnarray}
	\vev{2_14_1} [2_04_1][2_14_0][3_04_0], \quad \vev{2_14_1} [2_04_0][2_14_1][3_04_0], \quad
	\vev{2_14_1} [2_04_0][2_14_0][3_04_1],
\end{eqnarray}
which correspond to the following 3 operators
\begin{eqnarray}\label{eq:egRetrive}
    \phi_1 D^{\alpha}{}_{\dot{\alpha}} \psi^{\dagger}_{2\dot{\beta}} \psi^{\dagger}_{3\dot{\gamma}} D_{\alpha}{}^{\dot{\beta}} F_{\rm R 4}{}^{\dot{\alpha}\dot{\gamma}}, \quad \phi_1 D^{\alpha}{}_{\dot{\alpha}} \psi^{\dagger}_{2\dot{\beta}} \psi^{\dagger}_{3\dot{\gamma}} D_{\alpha}{}^{\dot{\alpha}} F_{\rm R 4}{}^{\dot{\beta}\dot{\gamma}}, \quad \phi_1 D^{\alpha}{}_{\dot{\alpha}} \psi^{\dagger}_{2\dot{\beta}} \psi^{\dagger}_{3\dot{\gamma}} D_{\alpha}{}^{\dot{\gamma}} F_{\rm R 4}{}^{\dot{\beta}\dot{\alpha}}.
\end{eqnarray}
It is straightforward to check that the 3 amplitudes only differ by the EOM of the $F_{\rm R 4}$, for example,
\eq{
	\vev{2_14_1} [2_04_1][2_14_0][3_04_0] - \vev{2_14_1} [2_04_0][2_14_1][3_04_0] = -\vev{2_14_1} [2_02_1][4_04_1][3_04_0],
}
or written in operators as,
\eq{
	\phi_1 D^{\alpha}{}_{\dot{\alpha}} \psi^{\dagger}_{2\dot{\beta}} \psi^{\dagger}_{3\dot{\gamma}} D_{\alpha}{}^{\dot{\beta}} F_{\rm R 4}{}^{\dot{\alpha}\dot{\gamma}} - \phi_1 D^{\alpha}{}_{\dot{\alpha}} \psi^{\dagger}_{2\dot{\beta}} \psi^{\dagger}_{3\dot{\gamma}} D_{\alpha}{}^{\dot{\alpha}} F_{\rm R 4}{}^{\dot{\beta}\dot{\gamma}} = -\phi_1 D^{\alpha\dot{\alpha}} \psi^{\dagger}_{2\dot{\alpha}} \psi^{\dagger}_{3\dot{\gamma}} D_{\alpha}{}^{\dot{\beta}} F_{\rm R 4\dot{\beta}}{}^{\dot{\gamma}},
}
thus we can choose any one of the three operators in (\ref{eq:egRetrive}) as the independent operator.

Here are some comments for the above method of obtaining the
off-shell basis. First, the method can be directly generalized to the cases where some fields are on-shell while others are off-shell in an EFT. Second, the method can be applied to the case where all fields are off-shell, but can not be applied to the case where all fields are on-shell. For instance, in the Green's basis in the SMEFT, if a class only involves $F_{\rm R}$s and derivatives, all fields in the class are taken to be on-shell and the previous method of removing the IBP eq.~(\ref{eq:IBP}) can not remove the IBP completely. In fact, the operator bases in that case are just the on-shell physical bases and can be obtained through the on-shell construction~\cite{Li:2020gnx,Li:2020xlh,Li:2022tec}.
 
\subsection{Off-shell Lorentz M-basis}\label{sec:lormbasis}

The method to obtain the off-shell y-basis in a class in section~\ref{sec:lorybasis} implies an algorithm to reduce any amplitude in the class into the basis. The algorithm is described as follows.
\begin{enumerate}
    \item Iteratively use the IBP relation eq.~(\ref{eq:IBP}) to move all derivatives on the first field to other fields.
    \item If the class involves $F_R$s, take the $F_R$s on-shell. Specifically, all $\lambda_{i,d_i}/\tilde\lambda_{i,d_i}$s of the derivatives acting on each of the $F_R$s are taken to be $\lambda_{i,0}$.
    \item Symmetrize the spinor indices corresponding to the derivatives acting on each field which is not $F_R$ using Young symmetrizers.
    \item Apply the Schouten identity eq.~(\ref{eq:Schouten}) to the symmetrized amplitude.
\end{enumerate}

With this algorithm, we can introduce the so-called off-shell m-basis where all basis vectors are monomials. In each class, the off-shell m-basis is connected with the off-shell y-basis by an invertible matrix of linear transformations $\mc{K}^{my}_{\mc{B}}$,
\eq{
        \mc{B}^m_i = \sum_j(\mc{K}^{my}_{\mc{B}})_{ij}\mc{B}^y_j.
    }
Furthermore, the algorithm can be applied to reduce an operator to the off-shell m-basis by decomposing the operator into the off-shell y-basis and utilizing the matrix $\mc{K}^{my}_{\mc{B}}$,
\eq{
        \mc{B} = \sum_j \mc{K}^y_j \mc{B}^y_j = \sum_{i,j} \mc{K}^y_j (\mc{K}^{my,-1})^{ji} \mc{B}^m_i \equiv \sum_i \mc{K}_{\mc{B},i} \mc{B}^m_i.
    }

For example, the following off-shell m-basis can be obtained to further simplify the result eq.~(\ref{eq:ybasis1}),
\eq{\label{eq:egmbasis}
    \mc{B}^{\text{(off-shell) $m$}}_{F_{\rm L} \psi^2 \phi D^2,1}&=F_{\rm L1}{}^{\mu\nu}(\psi_2 D_{\mu} \psi_3) D_{\nu} \phi_4, \\ \mc{B}^{\text{(off-shell) $m$}}_{F_{\rm L} \psi^2 \phi D^2,2}&=F_{\rm L1}{}_{\lambda}{}^{\nu}(\psi_2 i\sigma^{\lambda\mu} D_{\mu} \psi_3) D_{\nu} \phi_4, \\
    \mc{B}^{\text{(off-shell) $m$}}_{F_{\rm L} \psi^2 \phi D^2,3}&=F_{\rm L1}{}_{\lambda}{}^{\mu}(D_{\nu} D_{\mu} \psi_2 i\sigma^{\lambda\nu} \psi_3) \phi_4, \\ 
    \mc{B}^{\text{(off-shell) $m$}}_{F_{\rm L} \psi^2 \phi D^2,4}&=F_{\rm L1}{}^{\mu\nu}(\psi_2 i\sigma_{\mu\nu} D^{\lambda} D_{\lambda} \psi_3) \phi_4, \\
    \mc{B}^{\text{(off-shell) $m$}}_{F_{\rm L} \psi^2 \phi D^2,5}&=F_{\rm L1}{}^{\mu\nu}(\psi_2 i\sigma_{\mu\nu} \psi_3) D^{\lambda} D_{\lambda} \phi_4, \\
    \mc{B}^{\text{(off-shell) $m$}}_{F_{\rm L} \psi^2 \phi D^2,6}&=F_{\rm L1}{}_{\lambda}{}^{\mu}(D_{\mu} \psi_2 i\sigma^{\lambda\nu} D_{\nu} \psi_3) \phi_4, \\
    \mc{B}^{\text{(off-shell) $m$}}_{F_{\rm L} \psi^2 \phi D^2,7}&=F_{\rm L1}{}^{\mu\nu}(D_{\mu} \psi_2 D_{\nu} \psi_3) \phi_4, \\
    \mc{B}^{\text{(off-shell) $m$}}_{F_{\rm L} \psi^2 \phi D^2,8}&=F_{\rm L1}{}^{\mu\nu}(D_{\lambda} \psi_2 i\sigma_{\mu\nu} D^{\lambda} \psi_3) \phi_4, \\
    \mc{B}^{\text{(off-shell) $m$}}_{F_{\rm L} \psi^2 \phi D^2,9}&=F_{\rm L1}{}_{\lambda}{}^{\mu}(D_{\mu} \psi_2 i\sigma^{\lambda\nu} \psi_3) D_{\nu} \phi_4, \\
    \mc{B}^{\text{(off-shell) $m$}}_{F_{\rm L} \psi^2 \phi D^2,10}&=F_{\rm L1}{}^{\mu\nu}(D_{\mu} \psi_2 \psi_3) D_{\nu} \phi_4, \\
    \mc{B}^{\text{(off-shell) $m$}}_{F_{\rm L} \psi^2 \phi D^2,11}&=F_{\rm L1}{}^{\mu\nu}(D_{\lambda} \psi_2 i\sigma_{\mu\nu} \psi_3) D^{\lambda} \phi_4, \\
    \mc{B}^{\text{(off-shell) $m$}}_{F_{\rm L} \psi^2 \phi D^2,12}&=F_{\rm L1}{}^{\mu\nu}(\psi_2 i\sigma_{\mu\nu} D_{\lambda} \psi_3) D^{\lambda} \phi_4.
}
We can map each operator in eq~(\ref{eq:egmbasis}) to an amplitude and utilize the algorithm above to reduce the amplitude into the off-shell y-basis eq.~(\ref{eq:ybasis1}). The conversion matrix between the off-shell m-basis and the off-shell y-basis is given as
\begin{equation}
    \left(
\begin{array}{c}
 \mc{B}^{\text{(off-shell) $m$}}_{F_{\rm L} \psi^2 \phi D^2,1}\\
 \mc{B}^{\text{(off-shell) $m$}}_{F_{\rm L} \psi^2 \phi D^2,2}\\
 \mc{B}^{\text{(off-shell) $m$}}_{F_{\rm L} \psi^2 \phi D^2,3}\\
 \mc{B}^{\text{(off-shell) $m$}}_{F_{\rm L} \psi^2 \phi D^2,4}\\
 \mc{B}^{\text{(off-shell) $m$}}_{F_{\rm L} \psi^2 \phi D^2,5}\\
 \mc{B}^{\text{(off-shell) $m$}}_{F_{\rm L} \psi^2 \phi D^2,6}\\
 \mc{B}^{\text{(off-shell) $m$}}_{F_{\rm L} \psi^2 \phi D^2,7}\\
 \mc{B}^{\text{(off-shell) $m$}}_{F_{\rm L} \psi^2 \phi D^2,8}\\
 \mc{B}^{\text{(off-shell) $m$}}_{F_{\rm L} \psi^2 \phi D^2,9}\\
 \mc{B}^{\text{(off-shell) $m$}}_{F_{\rm L} \psi^2 \phi D^2,10}\\
 \mc{B}^{\text{(off-shell) $m$}}_{F_{\rm L} \psi^2 \phi D^2,11}\\
 \mc{B}^{\text{(off-shell) $m$}}_{F_{\rm L} \psi^2 \phi D^2,12}\\
\end{array}
\right) = \left(
\begin{array}{cccccccccccc}
 0 & 0 & 0 & 0 & 0 & 0 & 0 & 0 & 0 & 0 & -\frac{1}{4} & \frac{1}{4} \\
 0 & 0 & 0 & 0 & 0 & 0 & 0 & 0 & 0 & -\frac{1}{2} & \frac{1}{4} & \frac{1}{4} \\
 \frac{1}{2} & 0 & 0 & 0 & 0 & 0 & 0 & 0 & 0 & 0 & 0 & 0 \\
 0 & -1 & 0 & 0 & 0 & 0 & 0 & 0 & 0 & 0 & 0 & 0 \\
 0 & 0 & -1 & 0 & 0 & 0 & 0 & 0 & 0 & 0 & 0 & 0 \\
 0 & 0 & 0 & -\frac{1}{4} & 0 & -\frac{1}{4} & 0 & 0 & 0 & 0 & 0 & 0 \\
 0 & 0 & 0 & -\frac{1}{4} & 0 & \frac{1}{4} & 0 & 0 & 0 & 0 & 0 & 0 \\
 0 & 0 & 0 & 0 & -1 & 0 & 0 & 0 & 0 & 0 & 0 & 0 \\
 0 & 0 & 0 & 0 & 0 & 0 & -\frac{1}{4} & 0 & -\frac{1}{4} & 0 & 0 & 0 \\
 0 & 0 & 0 & 0 & 0 & 0 & -\frac{1}{4} & 0 & \frac{1}{4} & 0 & 0 & 0 \\
 0 & 0 & 0 & 0 & 0 & 0 & 0 & -1 & 0 & 0 & 0 & 0 \\
 0 & 0 & 0 & 0 & 0 & 0 & 0 & 0 & 0 & -1 & 0 & 0 \\
\end{array}
\right) \left(
\begin{array}{c}
 \mc{B}^{\text{(off-shell) $y$}}_{F_{\rm L} \psi^2 \phi D^2,1}\\
 \mc{B}^{\text{(off-shell) $y$}}_{F_{\rm L} \psi^2 \phi D^2,2}\\
 \mc{B}^{\text{(off-shell) $y$}}_{F_{\rm L} \psi^2 \phi D^2,3}\\
 \mc{B}^{\text{(off-shell) $y$}}_{F_{\rm L} \psi^2 \phi D^2,4}\\
 \mc{B}^{\text{(off-shell) $y$}}_{F_{\rm L} \psi^2 \phi D^2,5}\\
 \mc{B}^{\text{(off-shell) $y$}}_{F_{\rm L} \psi^2 \phi D^2,6}\\
 \mc{B}^{\text{(off-shell) $y$}}_{F_{\rm L} \psi^2 \phi D^2,7}\\
 \mc{B}^{\text{(off-shell) $y$}}_{F_{\rm L} \psi^2 \phi D^2,8}\\
 \mc{B}^{\text{(off-shell) $y$}}_{F_{\rm L} \psi^2 \phi D^2,9}\\
 \mc{B}^{\text{(off-shell) $y$}}_{F_{\rm L} \psi^2 \phi D^2,10}\\
 \mc{B}^{\text{(off-shell) $y$}}_{F_{\rm L} \psi^2 \phi D^2,11}\\
 \mc{B}^{\text{(off-shell) $y$}}_{F_{\rm L} \psi^2 \phi D^2,12}\\
\end{array}
\right).
\end{equation}
It is straightforward to check the above matrix is full-rank and thus invertible.

\subsection{Gauge Y-basis and M-basis}\label{sec:gauymbasis}

In the SMEFT, the non-Abelian gauge groups are the $SU(3)_C$ and the $SU(2)_W$.
The independent gauge factors of the gauge groups in the Green's basis are exactly the same as these in the on-shell basis, so we will just briefly review the method to construct the gauge y-basis and m-basis, and refer the readers to our previous works~\cite{Li:2020gnx,Li:2020xlh,Li:2022tec} for details.

In our previous works, we proposed a method to find all independent gauge factors of an operator as Levi-Civita tensors provided that all fields in the operator are expressed with the fundamental indices of the gauge groups.
The non-fundamental irreducible representations can be presented by multiple fundamental indices with particular symmetries. As the non-fundamental irreducible representations under the gauge groups of fields in the SM only include anti-fundamental and adjoint representations, the conversions between them and the fundamental irreducible representations are listed below:
\eq{\label{eq:example_T_gluon}
	&\epsilon_{acd}\lambda^A{}_b^d G^A = G_{abc} \sim \young(ab,c),\\
	&\epsilon_{abc}Q^{\dagger,{c}} = Q^\dagger_{ab} \sim \young(a,b),\\
	&\epsilon_{jk}\tau^I{}_i^k W^I = W_{ij}\sim \young(ij) ,\\
	&\epsilon_{ij}H^{\dagger, j} = H_i \sim \young(i) .
}
Then the $y$-basis of group factors are constructed by applying the generalized Littlewood-Richardson rules to the corresponding Young tableaux of all fields in the operator. Each unique way to obtain a singlet Young tableau from the Young tableaux of all fields gives a independent gauge factor.
The singlet Young tableaux of the $SU(2)_W$ and the $SU(3)_C$ groups are as following:
\begin{eqnarray}
	SU(3)_C:\ \underbrace{\yng(1,1,1)\ ...\ \yng(1,1,1)}_{\let\scriptstyle\textstyle\text{\large $n_{\rm box}/3$}},\quad
	SU(2)_W:\ \underbrace{\yng(1,1)\ ...\ \yng(1,1)}_{\let\scriptstyle\textstyle\text{\large $n_{\rm box}/2$}},
\end{eqnarray}
where $n_{\rm box}$ is the total number of boxes in the Young tableau, which equals to the total number of fundamental indices of the fields in an operator.

For the gauge factors, a basis consisted of monomial gauge factors can be constructed similarly as the Lorentz m-basis, called the gauge m-basis. In Ref.~\cite{Li:2022tec}, an algorithm to select $N$ independent gauge factors from a set of over-complete candidate gauge factors $\{T_m|m \in \{1,2,\cdots,M\}\}$ is introduced. The main idea is to start by introducing the inner product defined between two group factors of the same rank
\begin{equation}
    (T_i, T_j)=\sum_{\{a_i\}}(T_i^{a_1a_2\dots})^*T_j^{a_1a_2\dots}.
\end{equation}
With this inner product, one can iteratively determine the independence of a new candidate gauge factor $T_m$ with a set of gauge factors $\{T'_i|i \in \{1,2,\cdots,n\}\}$ that have been checked to be independent by adding the $T_m$ to the set to be $T'_{n+1}$ and checking if the metric $g_{ij}=(T'_i, T'_j)$, where $i,j \in \{1,2,\cdots,n+1\}$, is invertible. If yes, $T'_{n+1}$ is kept to be the $(n+1)$th independent gauge factor, otherwise $T'_{n+1}$ is 
dropped, and then one move to the next candidate gauge factor $T_{m+1}$. One can repeat the above steps until the number of independent gauge factors reaches the upper limit $N$. The upper limit $N$ is the dimension of the corresponding gauge y-basis, and the set of over-complete candidate gauge factors can be obtained by selecting monomial gauge factors from the corresponding gauge y-basis.

We can also use the metric introduced above to find the projection of a given gauge factor onto the m-basis by the inner product
\eq{\label{eq:}
    (T^m_i,T) = \sum_j \mc{K}_{T,j} (T^m_i,T^m_j) = \sum_j g_{ij} \mc{K}_{T,j}, \quad \Rightarrow \mc{K}_{T,i} = \sum_j g^{ij} (T^m_j,T),
}
where $g^{ij}$ is the inverse metric $g^{ij}g_{jk} = \delta^i_{\ k}$, and the invertibility is guaranteed by the independence of the m-basis $\{T^m_i\}$.

\subsection{Off-shell P-basis and F-basis}\label{sec:pbasis}

During the construction of the off-shell y-basis and m-basis, fields in an operator are regarded as distinguishable and sorted in a certain order. However, an actual operator could involve repeated fields, and the operator should be totally symmetric under any permutation of bosonic fields while totally antisymmetric under any permutation of fermionic fields due to the spin-statistics. The spin-statistics constraint is trivial for distinguishable fields, but can be tricky for repeated fields, since any permutation of the repeated fields is equivalent to the permutation of all indices in these repeated fields. The spin-statistics constraint can be written as
\eq{\label{eq:exchpar}
	\mc{O}\big( \underbrace{\phi^{a_1},\dots,\phi^{a_m}}_{m},\dots \big) 
	= \mc{R}_{\phi}(\pi)\mc{O}
	\big( \underbrace{\phi^{a_{\pi(1)}},\dots,\phi^{a_{\pi(m)}}}_{m},\dots \big), \\
	\mc{R}_{\phi}(\pi) = \left\{\begin{array}{ll} 1 & \text{bosonic } \phi \\ (-1)^{{\rm sgn}(\pi)} & \text{fermionic } \phi \end{array} \right. ,\qquad \pi\in S_m ,
}
where $\mc{R}_{\phi}(\pi)$ denote the representation of the permutation $\pi$ of the identical particles $\phi$, and $(-1)^{\rm sgn}(\pi)$ is the signature of the permutation $\pi$. To remove the redundancy of the constraint, we treat the repeated fields as flavor multiplets with different flavor indices, so that the constraint eq.~(\ref{eq:exchpar}) can be further explicated as
\begin{eqnarray}
\underbrace{\pi\circ {\cal O}^{\{f_{k},...\}}}_{\rm permute\ flavor} &=& \mc{R}_{\phi}(\pi^{-1})\underbrace{\left(\pi\circ T_{{\rm SU3}}^{\{g_k,...\}}\right)\left(\pi\circ T_{{\rm SU2}}^{\{h_k,...\}}\right)}_{\rm permute\ gauge}\underbrace{\left(\pi\circ{\cal M}^{\{f_k,...\}}_{\{g_{k},...\},\{h_{k},...\}}\right)}_{\rm permute\ Lorentz},
\label{eq:tperm}
\end{eqnarray}
where $f_{k}$, $g_k$ and $h_k$ are the flavor, the $SU(3)_C$ and the $SU(2)_W$ gauge group indices for different sets of repeated fields respectively. We can organize operators in the off-shell m-basis into irreducible representations of the symmetric group by applying the group algebra projectors $b^\lambda_i$~\cite{Li:2020xlh} to the flavor indices of the repeated fields, which is equivalent to permuting the gauge and Lorentz indices using eq.~(\ref{eq:tperm}). The operators obtained in this way are called p-basis. Furthermore, the operators in the p-basis are also flavor tensors of the corresponding $SU(n_f)$ group, and flavor tensors as different bases of the same irreducible representations of the symmetric group, span the same space of the $SU(n_f)$ group. Therefore we only keep one basis of each irreducible representation of the symmetric
group whose dimension is larger than one. After removing all redundancies of flavor, the operator basis is called the f-basis.

In the following, we take the type $Q^3 H e^{\dagger}_{_\mathbb{C}} D$ as an example to illustrate the method to dealing with the repeated fields and the process to obtaining the off-shell p-basis and the off-shell f-basis of the type. The off-shell Lorentz y-basis and m-basis of that type are
\begin{align}
    \begin{array}{c|c}
        \text{Lorentz Y-basis} &  \\
        \hline
        \mc{B}^{\text{(off-shell) $y$}}_{\psi^3 \phi \psi^{\dagger} D,1} & \vev{1_02_0}\vev{2_13_0}[2_15_0] \\
        \mc{B}^{\text{(off-shell) $y$}}_{\psi^3 \phi \psi^{\dagger} D,2} & \vev{1_02_1}\vev{2_03_0}[2_15_0] \\
        \mc{B}^{\text{(off-shell) $y$}}_{\psi^3 \phi \psi^{\dagger} D,3} & \vev{1_02_0}\vev{3_03_1}[3_15_0] \\
        \mc{B}^{\text{(off-shell) $y$}}_{\psi^3 \phi \psi^{\dagger} D,4} & \vev{1_03_0}\vev{2_03_1}[3_15_0] \\
        \mc{B}^{\text{(off-shell) $y$}}_{\psi^3 \phi \psi^{\dagger} D,5} & \vev{1_02_0}\vev{3_04_1}[4_15_0] \\
        \mc{B}^{\text{(off-shell) $y$}}_{\psi^3 \phi \psi^{\dagger} D,6} & \vev{1_03_0}\vev{2_04_1}[4_15_0] \\
        \mc{B}^{\text{(off-shell) $y$}}_{\psi^3 \phi \psi^{\dagger} D,7} & \vev{1_02_0}\vev{3_05_1}[5_05_1] \\
        \mc{B}^{\text{(off-shell) $y$}}_{\psi^3 \phi \psi^{\dagger} D,8} & \vev{1_03_0}\vev{2_05_1}[5_05_1]
    \end{array}, \quad
    \begin{array}{c|c}
        \text{Lorentz M-basis} &  \\
        \hline
        \mc{B}^{\text{(off-shell) $m$}}_{\psi^3 \phi \psi^{\dagger} D,1} & (\psi_1\psi_2) (\psi_3 i\sigma_{\mu} \psi^{\dagger}_5) D^{\mu} \phi_4 \\
        \mc{B}^{\text{(off-shell) $m$}}_{\psi^3 \phi \psi^{\dagger} D,2} & (\psi_1\psi_3) (\psi_2 i\sigma_{\mu} \psi^{\dagger}_5) D^{\mu} \phi_4 \\
        \mc{B}^{\text{(off-shell) $m$}}_{\psi^3 \phi \psi^{\dagger} D,3} & (\psi_1 D^{\mu} \psi_2) (\psi_3 i\sigma_{\mu} \psi^{\dagger}_5) \phi_4 \\
        \mc{B}^{\text{(off-shell) $m$}}_{\psi^3 \phi \psi^{\dagger} D,4} & (D^{\mu} \psi_2 \psi_3) (\psi_1 i\sigma_{\mu} \psi^{\dagger}_5) \phi_4 \\
        \mc{B}^{\text{(off-shell) $m$}}_{\psi^3 \phi \psi^{\dagger} D,5} & (\psi_1\psi_2) (D^{\mu} \psi_3 i\sigma_{\mu} \psi^{\dagger}_5) \phi_4 \\
        \mc{B}^{\text{(off-shell) $m$}}_{\psi^3 \phi \psi^{\dagger} D,6} & (\psi_1 D^{\mu} \psi_3) (\psi_2 i\sigma_{\mu} \psi^{\dagger}_5) \phi_4 \\
        \mc{B}^{\text{(off-shell) $m$}}_{\psi^3 \phi \psi^{\dagger} D,7} & (\psi_1\psi_2) (\psi_3 i\sigma_{\mu} D^{\mu} \psi^{\dagger}_5) \phi_4 \\
        \mc{B}^{\text{(off-shell) $m$}}_{\psi^3 \phi \psi^{\dagger} D,8} & (\psi_1 \psi_3) (\psi_2 i\sigma_{\mu} D^{\mu} \psi^{\dagger}_5) \phi_4
    \end{array},
\end{align}
with the $\mc{K}^{my}_{\mc{B}}$ is given as
\begin{eqnarray}
    \mc{K}^{my}_{\mc{B}} = \left(
\begin{array}{cccccccc}
 0 & 0 & 0 & 0 & 1 & 0 & 0 & 0 \\
 0 & 0 & 0 & 0 & 0 & -1 & 0 & 0 \\
 -1 & 0 & 0 & 0 & 0 & 0 & 0 & 0 \\
 0 & 1 & 0 & 0 & 0 & 0 & 0 & 0 \\
 0 & 0 & 1 & 0 & 0 & 0 & 0 & 0 \\
 0 & 0 & 0 & -1 & 0 & 0 & 0 & 0 \\
 0 & 0 & 0 & 0 & 0 & 0 & -1 & 0 \\
 0 & 0 & 0 & 0 & 0 & 0 & 0 & 1 \\
\end{array}
\right).
\end{eqnarray}
Permuting the three $Q$s in $Q^3 H e^{\dagger}_{_\mathbb{C}} D$ form a $S_3$ group. With the algorithm described in section~\ref{sec:lormbasis}, we can deduce the matrix representations of the generators of the $S_3$ group, $(1 \ 2)$ and $(1 \ 2 \ 3)$, under the off-shell Lorentz y-basis, as $\mc{R}_{\mc{B}}^y[(1 \ 2)]$ and $\mc{R}_{\mc{B}}^y[(1 \ 2 \ 3)]$. For example,
\eq{
    (1 \ 2) \ \mc{B}^{\text{(off-shell) $y$}}_{\psi^3 \phi \psi^{\dagger} D,1} &= \vev{2_01_0}\vev{1_13_0}[1_15_0] \\
    &= - \vev{2_01_0}\vev{2_13_0}[2_15_0] - \vev{2_01_0}\vev{3_13_0}[3_15_0] - \vev{2_01_0}\vev{4_13_0}[4_15_0] \\
    &\quad \,- \vev{2_01_0}\vev{5_13_0}[5_15_0] \\
    &= \mc{B}^{\text{(off-shell) $y$}}_{\psi^3 \phi \psi^{\dagger} D,1} - \mc{B}^{\text{(off-shell) $y$}}_{\psi^3 \phi \psi^{\dagger} D,3} - \mc{B}^{\text{(off-shell) $y$}}_{\psi^3 \phi \psi^{\dagger} D,5} + \mc{B}^{\text{(off-shell) $y$}}_{\psi^3 \phi \psi^{\dagger} D,7},
}
where we used the IBP relation eq.~(\ref{eq:IBP}) in the second equality. After deducing the $\mc{R}_{\mc{B}}^y[(1 \ 2)]$ and the $\mc{R}_{\mc{B}}^y[(1 \ 2 \ 3)]$, the matrix representations of those generators under the off-shell Lorentz m-basis can be obtained by $\mc{R}^{m}[\pi] = \mc{R}_{\phi}[\pi^{-1}] \mc{K}^{my} \mc{R}^{y}[\pi] (\mc{K}^{my})^{-1}$, where the matrix representation of the permutation of fields, $\mc{R}_{\phi}[\pi^{-1}]$, is taken into account here.
The matrices are given as follows,
\begin{eqnarray}\label{eq:mrepLor1}
    \mc{R}_{\mc{B}}^m[(1 \ 2)]= -\mc{K}^{my}_{\mc{B}} \mc{R}_{\mc{B}}^{y}[(1 \ 2)] (\mc{K}_{\mc{B}}^{my})^{-1}=\left(
\begin{array}{cccccccc}
 1 & 0 & 0 & 0 & 0 & 0 & 0 & 0 \\
 -1 & -1 & 0 & 0 & 0 & 0 & 0 & 0 \\
 -1 & 0 & -1 & 0 & -1 & 0 & -1 & 0 \\
 0 & -1 & 1 & 1 & 0 & -1 & 0 & -1 \\
 0 & 0 & 0 & 0 & 1 & 0 & 0 & 0 \\
 0 & 0 & 0 & 0 & -1 & -1 & 0 & 0 \\
 0 & 0 & 0 & 0 & 0 & 0 & 1 & 0 \\
 0 & 0 & 0 & 0 & 0 & 0 & -1 & -1 \\
\end{array}
\right), \\ \mc{R}_{\mc{B}}^m[(1 \ 2 \ 3)]= \mc{K}^{my}_{\mc{B}} \mc{R}_{\mc{B}}^{y}[(1 \ 2 \ 3)] (\mc{K}_{\mc{B}}^{my})^{-1}=\left(
\begin{array}{cccccccc}
 -1 & -1 & 0 & 0 & 0 & 0 & 0 & 0 \\
 1 & 0 & 0 & 0 & 0 & 0 & 0 & 0 \\
 0 & 0 & 0 & 0 & -1 & -1 & 0 & 0 \\
 0 & 0 & 0 & 0 & 0 & 1 & 0 & 0 \\
 1 & 1 & 0 & -1 & 1 & 1 & 1 & 1 \\
 -1 & 0 & -1 & 0 & -1 & 0 & -1 & 0 \\
 0 & 0 & 0 & 0 & 0 & 0 & -1 & -1 \\
 0 & 0 & 0 & 0 & 0 & 0 & 1 & 0 \\
\end{array}
\right).\label{eq:mrepLor2}
\end{eqnarray}

The gauge m-basis of the type $Q^3 H e^{\dagger}_{_\mathbb{C}} D$ is
\begin{align}
    \begin{array}{c|c}
        \text{Gauge M-basis} &  \\
        \hline
        T^m_{Q^3 H e^{\dagger}_{_\mathbb{C}} D,1} & \epsilon^{a_1a_2a_3} \epsilon^{i_1i_3} \epsilon^{i_2i_4} \\
        T^m_{Q^3 H e^{\dagger}_{_\mathbb{C}} D,2} & \epsilon^{a_1a_2a_3} \epsilon^{i_1i_2} \epsilon^{i_3i_4}
    \end{array}
\end{align}
Similarly, we can find the matrix representation of the generators under the above gauge m-basis with the method introduced in section~\ref{sec:gauymbasis},
\begin{eqnarray}\label{eq:mrepgau}
    \mc{D}^m_{T}[(1 \ 2)]=\left(
\begin{array}{ccc}
 -1 & 1 \\
 0 & 1
\end{array}
\right), \quad \mc{D}^m_{T}[(1 \ 2 \ 3)]=\left(
\begin{array}{ccc}
 0 & -1 \\
 1 & -1
\end{array}
\right).
\end{eqnarray}

The full m-basis should be the direct product of the gauge m-basis and the off-shell Lorentz m-basis, and the matrix representations of the $S_3$ generators under the full m-basis of the type $Q^3 H e^{\dagger}_{_\mathbb{C}} D$ should be the direct product of that under the gauge m-basis and the off-shell Lorentz m-basis from eq.~(\ref{eq:mrepgau}), eq.~(\ref{eq:mrepLor1}) and eq.~(\ref{eq:mrepLor2}),
\begin{eqnarray}
    \mc{D}^m(\pi) = \mc{D}^m_{T}(\pi) \otimes \mc{D}^m_{\mc{B}}(\pi).
\end{eqnarray}
The matrix representation of the $S_3$ generators can be used to generate the matrix representation of the group algebra projectors $b^\lambda_i$ which would project out the part of a tensor that is the irreducible representation $\lambda$ of the symmetric group when acting on the tensor. For instance, there are two group algebra projectors of the $[2,1]$ representation, $b^{[2,1]}_1$ and $b^{[2,1]}_2$, since the representation is 2-dimensional, and the matrix representation of the first group algebra projector is
\begin{eqnarray}
    b^{[2,1]}_1 = \left(
\begin{array}{cccccccccccccccc}
 \frac{1}{3} & \frac{2}{3} & 0 & 0 & 0 & 0 & 0 & 0 & \frac{1}{3} & -\frac{1}{3} & 0 & 0 & 0 & 0 & 0 & 0 \\
 0 & 0 & 0 & 0 & 0 & 0 & 0 & 0 & 0 & 0 & 0 & 0 & 0 & 0 & 0 & 0 \\
 \frac{1}{3} & -\frac{1}{3} & 1 & 0 & \frac{1}{3} & -\frac{1}{3} & \frac{1}{3} & -\frac{1}{3} & -\frac{1}{3} & \frac{1}{3} & -\frac{2}{3} &
   -\frac{1}{3} & -\frac{1}{3} & \frac{1}{3} & -\frac{1}{3} & \frac{1}{3} \\
 -\frac{1}{3} & \frac{1}{3} & -1 & 0 & -\frac{1}{3} & \frac{1}{3} & -\frac{1}{3} & \frac{1}{3} & \frac{1}{3} & -\frac{1}{3} & \frac{2}{3} & \frac{1}{3}
   & \frac{1}{3} & -\frac{1}{3} & \frac{1}{3} & -\frac{1}{3} \\
 -\frac{1}{3} & -\frac{1}{3} & -\frac{1}{3} & 0 & -\frac{1}{3} & -\frac{1}{3} & -\frac{1}{3} & -\frac{1}{3} & 0 & 0 & \frac{1}{3} & \frac{1}{3} &
   \frac{1}{3} & 0 & 0 & 0 \\
 0 & \frac{1}{3} & -\frac{1}{3} & 0 & \frac{1}{3} & 1 & 0 & \frac{1}{3} & 0 & 0 & 0 & -\frac{1}{3} & -\frac{1}{3} & -\frac{1}{3} & 0 & 0 \\
 0 & 0 & 0 & 0 & 0 & 0 & \frac{1}{3} & \frac{2}{3} & 0 & 0 & 0 & 0 & 0 & 0 & \frac{1}{3} & -\frac{1}{3} \\
 0 & 0 & 0 & 0 & 0 & 0 & 0 & 0 & 0 & 0 & 0 & 0 & 0 & 0 & 0 & 0 \\
 \frac{1}{3} & \frac{2}{3} & 0 & 0 & 0 & 0 & 0 & 0 & \frac{1}{3} & -\frac{1}{3} & 0 & 0 & 0 & 0 & 0 & 0 \\
 -\frac{1}{3} & -\frac{2}{3} & 0 & 0 & 0 & 0 & 0 & 0 & -\frac{1}{3} & \frac{1}{3} & 0 & 0 & 0 & 0 & 0 & 0 \\
 0 & -\frac{1}{3} & \frac{1}{3} & 0 & 0 & -\frac{1}{3} & 0 & -\frac{1}{3} & -\frac{1}{3} & 0 & 0 & \frac{1}{3} & -\frac{1}{3} & 0 & -\frac{1}{3} & 0 \\
 -\frac{1}{3} & 0 & -\frac{2}{3} & 0 & -\frac{1}{3} & 0 & -\frac{1}{3} & 0 & 0 & -\frac{1}{3} & \frac{2}{3} & \frac{2}{3} & 0 & -\frac{1}{3} & 0 &
   -\frac{1}{3} \\
 -\frac{1}{3} & -\frac{1}{3} & -\frac{1}{3} & 0 & -\frac{1}{3} & -\frac{1}{3} & -\frac{1}{3} & -\frac{1}{3} & \frac{1}{3} & \frac{1}{3} & 0 &
   -\frac{1}{3} & 1 & \frac{1}{3} & \frac{1}{3} & \frac{1}{3} \\
 0 & \frac{1}{3} & -\frac{1}{3} & 0 & 0 & \frac{1}{3} & 0 & \frac{1}{3} & 0 & -\frac{1}{3} & \frac{1}{3} & \frac{1}{3} & -\frac{1}{3} & -\frac{1}{3} &
   0 & -\frac{1}{3} \\
 0 & 0 & 0 & 0 & 0 & 0 & \frac{1}{3} & \frac{2}{3} & 0 & 0 & 0 & 0 & 0 & 0 & \frac{1}{3} & -\frac{1}{3} \\
 0 & 0 & 0 & 0 & 0 & 0 & -\frac{1}{3} & -\frac{2}{3} & 0 & 0 & 0 & 0 & 0 & 0 & -\frac{1}{3} & \frac{1}{3} \\
\end{array}
\right),
\end{eqnarray}
where
\begin{eqnarray}
    b^{[2,1]}_1 = \mc{Y}^{[2,1]}_1 = \dfrac{1}{3} \left(E+(1 \ 2) - (1 \ 3) - (1 \ 3 \ 2)\right).
\end{eqnarray}
The matrix is rank-5, and we can choose the operators corresponding to the 1, 3, 5, 6 and 11 rows of the matrix as the independent operators that are of the first group algebra projector $b^{[2,1]}_1$ of the $[2,1]$ representation,
\eq{\label{eq:egopes}
    \Bl{\mc{Y}\left[\tiny{\young(pr,s)}\right]i \epsilon {}^{abc}\epsilon {}^{ik}\epsilon {}^{jl} D^{\mu } H_l{}  ( Q_s{}_{ck}{} \sigma_{\mu }{} e_{_\mathbb{C}}{}^{\dagger} _t ) ( Q_p{}_{ai}{} Q_r{}_{bj}{} )}, \\
    \Bl{\mc{Y}\left[\tiny{\young(pr,s)}\right]i \epsilon {}^{abc}\epsilon {}^{ik}\epsilon {}^{jl}H_l{} ( Q_s{}_{ck}{} \sigma_{\mu }{} e_{_\mathbb{C}}{}^{\dagger} _t ) ( Q_p{}_{ai}{} D^{\mu } Q_r{}_{bj}{}  )}, \\
    \mc{Y}\left[\tiny{\young(pr,s)}\right]i \epsilon {}^{abc}\epsilon {}^{ik}\epsilon {}^{jl}H_l{} ( D^{\mu } Q_s{}_{ck}{}  \sigma_{\mu }{} e_{_\mathbb{C}}{}^{\dagger} _t ) ( Q_p{}_{ai}{} Q_r{}_{bj}{} ), \\
    \mc{Y}\left[\tiny{\young(pr,s)}\right]i \epsilon {}^{abc}\epsilon {}^{ik}\epsilon {}^{jl}H_l{} ( Q_r{}_{bj}{} \sigma_{\mu }{} e_{_\mathbb{C}}{}^{\dagger} _t ) ( Q_p{}_{ai}{} D^{\mu } Q_s{}_{ck}{}  ), \\
    \mc{Y}\left[\tiny{\young(pr,s)}\right]i \epsilon {}^{abc}\epsilon {}^{ij}\epsilon {}^{kl}H_l{} ( Q_s{}_{ck}{} \sigma_{\mu }{} e_{_\mathbb{C}}{}^{\dagger} _t ) ( Q_p{}_{ai}{} D^{\mu } Q_r{}_{bj}{}  ).
}
In eq.~(\ref{eq:egopes}), we utilize the method introduced in Ref.~\cite{Li:2022tec} to reduce an over-complete basis into the on-shell basis, and distinguish operators that are in the on-shell basis, which are the operators in blue, from operators that only exist in the Green's basis, which are the operators in black. The choice of determining which operators are basis vectors of the on-shell basis is not unique.

The $SU(n_f)$ space spanned by flavor tensors of the second group algebra projector of the $[2,1]$ representation, $b^{[2,1]}_2$, is the same as that spanned by flavor tensors of $b^{[2,1]}_1$, as we elaborated in Ref.~\cite{Li:2022tec}, so we only keep the space spanned by flavor tensors of $b^{[2,1]}_1$. Considering the group algebra projectors of all 3 irreducible representations of the $S_3$ group, $[3]$, $[2,1]$ and $[1,1,1]$, the complete off-shell f-basis of the type $Q^3 H e^{\dagger}_{_\mathbb{C}} D$ is
\begin{eqnarray}
    \begin{array}{c|l}
		
		\multirow{6}*{$\mathcal{O}_{Q^3 H e^{\dagger}_{_\mathbb{C}} D}^{\left(1\sim11\right)}$}
		
		&\Bl{\mathcal{Y}\left[\tiny{\young(prs)}\right]i \epsilon {}^{abc}\epsilon {}^{ik}\epsilon {}^{jl} D^{\mu } H_l{}  ( Q_s{}_{ck}{} \sigma_{\mu }{} e_{_\mathbb{C}}{}^{\dagger} _t{}{}{} ) ( Q_p{}_{ai}{} Q_r{}_{bj}{} )},
		
		\Bl{\mc{Y}\left[\tiny{\young(pr,s)}\right]i \epsilon {}^{abc}\epsilon {}^{ik}\epsilon {}^{jl} D^{\mu } H_l{}  ( Q_s{}_{ck}{} \sigma_{\mu }{} e_{_\mathbb{C}}{}^{\dagger} _t ) ( Q_p{}_{ai}{} Q_r{}_{bj}{} )},
		
		\\&\Bl{\mc{Y}\left[\tiny{\young(pr,s)}\right]i \epsilon {}^{abc}\epsilon {}^{ik}\epsilon {}^{jl}H_l{} ( Q_s{}_{ck}{} \sigma_{\mu }{} e_{_\mathbb{C}}{}^{\dagger} _t ) ( Q_p{}_{ai}{} D^{\mu } Q_r{}_{bj}{}  )},
		
		\quad\Bl{\mathcal{Y}\left[\tiny{\young(p,r,s)}\right]i \epsilon {}^{abc}\epsilon {}^{ik}\epsilon {}^{jl} D^{\mu } H_l{}  ( Q_s{}_{ck}{} \sigma_{\mu }{} e_{_\mathbb{C}}{}^{\dagger} _t{}{}{} ) ( Q_p{}_{ai}{} Q_r{}_{bj}{} )},
		
		\\&\mathcal{Y}\left[\tiny{\young(prs)}\right]i \epsilon {}^{abc}\epsilon {}^{ik}\epsilon {}^{jl}H_l{} ( Q_s{}_{ck}{} \sigma_{\mu }{} e_{_\mathbb{C}}{}^{\dagger} _t{}{}{} ) ( Q_p{}_{ai}{} D^{\mu } Q_r{}_{bj}{}  ),
		
		\quad\mathcal{Y}\left[\tiny{\young(prs)}\right]i \epsilon {}^{abc}\epsilon {}^{ik}\epsilon {}^{jl}H_l{} ( Q_r{}_{bj}{} \sigma_{\mu }{} e_{_\mathbb{C}}{}^{\dagger} _t{}{}{} ) ( Q_p{}_{ai}{} D^{\mu } Q_s{}_{ck}{}  ),
		
		\\&\mathcal{Y}\left[\tiny{\young(pr,s)}\right]i \epsilon {}^{abc}\epsilon {}^{ik}\epsilon {}^{jl}H_l{} ( D^{\mu } Q_s{}_{ck}{}  \sigma_{\mu }{} e_{_\mathbb{C}}{}^{\dagger} _t{}{}{} ) ( Q_p{}_{ai}{} Q_r{}_{bj}{} ),
		
		\quad\mathcal{Y}\left[\tiny{\young(pr,s)}\right]i \epsilon {}^{abc}\epsilon {}^{ik}\epsilon {}^{jl}H_l{} ( Q_r{}_{bj}{} \sigma_{\mu }{} e_{_\mathbb{C}}{}^{\dagger} _t{}{}{} ) ( Q_p{}_{ai}{} D^{\mu } Q_s{}_{ck}{}  ),
		
		\\&\mathcal{Y}\left[\tiny{\young(pr,s)}\right]i \epsilon {}^{abc}\epsilon {}^{ij}\epsilon {}^{kl}H_l{} ( Q_s{}_{ck}{} \sigma_{\mu }{} e_{_\mathbb{C}}{}^{\dagger} _t{}{}{} ) ( Q_p{}_{ai}{} D^{\mu } Q_r{}_{bj}{}  ),
		
		\quad\mathcal{Y}\left[\tiny{\young(p,r,s)}\right]i \epsilon {}^{abc}\epsilon {}^{ik}\epsilon {}^{jl}H_l{} ( Q_s{}_{ck}{} \sigma_{\mu }{} e_{_\mathbb{C}}{}^{\dagger} _t{}{}{} ) ( Q_p{}_{ai}{} D^{\mu } Q_r{}_{bj}{}  ),
		
		\\&\mathcal{Y}\left[\tiny{\young(p,r,s)}\right]i \epsilon {}^{abc}\epsilon {}^{ik}\epsilon {}^{jl}H_l{} ( D^{\mu } Q_s{}_{ck}{}  \sigma_{\mu }{} e_{_\mathbb{C}}{}^{\dagger} _t{}{}{} ) ( Q_p{}_{ai}{} Q_r{}_{bj}{} ).
		
	\end{array}
\end{eqnarray}

\section{Dimension-8 Operators in the Green's Basis}\label{sec:Operators}

In this section, the complete and independent Green's basis of the dimension-8 operators in the SMEFT is listed. We also present the statistics of the dimension-8 Green's basis in the SMEFT in table~\ref{tab:statistic1} and table~\ref{tab:statistic2}, where the classes are slightly different from those in table~\ref{tab:classes8} since the conjugate classes of some classes are written explicitly. There are 4 classes in table~\ref{tab:classes8} vanishing in table~\ref{tab:statistic1} and table~\ref{tab:statistic2}, $F^2_{\rm R} D^4$, $F_{\rm L} F_{\rm R} D^4$, $F^3_{\rm R} D^2$ and $F_{\rm R} \phi^2 D^4$, due to the fact that $F_{\rm R}$ is taken to be on-shell in our convention.

In the following operator basis, the classes and types in blue are classes and types where no new operator shows up compared to those in the on-shell basis, and the operators in blue are operators that can be chosen to be in the on-shell basis. Of course, this choice is not unique. For a real type, the operators in the type are not Hermitian in general, and should combine their Hermitian conjugates to be Hermitian operators. For a complex type, the Hermitian conjugate of the type and the Hermitian conjugates of the operators in the type are omitted unless the type involves gauge bosons and derivatives.

The gauge bosons in the following operator basis are presented in the chiral basis as $F_{\rm L}/F_{\rm R}$, and the conversion rules between the chiral basis and the commonly used $F/\tilde{F}$ basis are given as
\begin{align}
F_{\rm{L}/\rm{R}}=\frac12\left(F\mp i\tilde{F} \right), \quad \tilde{F}^{\mu\nu}=\frac12\epsilon^{\mu\nu\rho\eta}F_{\rho\eta}.
\label{eq:FLR1}
\end{align}

\begin{table}[htbp]
	\centering	
	\begin{align}{\small
		\begin{array}{cc|c|c|c|c|c}
			\hline
			N & (n,\tilde{n}) & \text{Classes} & \mathcal{N}_{\text{type}} & \mathcal{N}_{\text{term}} & \mathcal{N}_{\text{operator}} & \text{Equations}\\
			\hline\hline
			2 & (4,2) & F^2_{\rm L} D^4 &  3 & 3 &3& (\ref{op:fl2d4f}-\ref{op:fl2d4l})\\
			\cline{2-7}
			& (3,3) & \phi^2 D^6 &  1 & 1 &1& (\ref{op:h2d6})\\
			& & \psi \psi^{\dagger}{} D^5 & 5 & 5 &5n_f^2& (\ref{op:ppgd5f}-\ref{op:ppgd5l})\\
			\hline
			3 & (4,1) & F^3_{\rm L} D^2 & 4 & 8 & 8 & (\ref{op:fl3d2f}-\ref{op:fl3d2l})\\
			\cline{2-7}
			 & (3,2) & F^2_{\rm L} F_{\rm R} D^2 & 4 & 4 & 4 & (\ref{op:fl2frd2f}-\ref{op:fl2frd2l})\\
			 & & F_{\rm L} \phi^2 D^4 & 2 & 6 & 6 & (\ref{op:flh2d4f}-\ref{op:flh2d4l})\\
			 & & F_{\rm L} \psi \psi^{\dagger}{} D^3 & 10 & 80 & 80n_f^2 & (\ref{op:flppgd3f}-\ref{op:flppgd3l})\\
			 & & \psi^2 \phi D^4 +h.c. & 6 & 54 & 54n_f^2 & (\ref{op:p2hd4f}-\ref{op:p2hd4l})\\
			\cline{2-7}
			 & (2,3) & F_{\rm L}F^2_{\rm R} D^2 & 4 & 4 & 4 & (\ref{op:flfr2d2f}-\ref{op:flfr2d2l})\\
			 & & F_{\rm R} \psi \psi^{\dagger}{} D^3 & 10 & 20 & 20n_f^2 & (\ref{op:frppgd3f}-\ref{op:frppgd3l})\\
			\hline
			4 & (4,0) & \Bl{F^4_{\rm L}+h.c.} &  \Bl{14} & \Bl{26} &\Bl{26}& (\ref{op:fl4f}-\ref{op:fl4l})\\
			\cline{2-7}
			 & (3,1) & F^2_{\rm L}\psi\psi^{\dagger}{}D & \Bl{11}+10 & \Bl{11}+94 &\Bl{11n^2_f}+94n^2_f& (\ref{op:fl2ppgdf}-\ref{op:fl2ppgdl})\\
			  & & \psi^4D^2+h.c. & \Bl{8} & \Bl{32}+192 &\Bl{n_f^3(17n_f-1)}+102n_f^4& (\ref{op:p4d2f}-\ref{op:p4d2l})\\
			  & & F_{\rm L}\psi^2\phi D^2 & \Bl{8} & \Bl{16}+80 &\Bl{16n^2_f}+80n_f^2& (\ref{op:flp2hd2f}-\ref{op:flp2hd2l})\\
			  & & F^2_{\rm L}\phi^2D^2 & \Bl{4} & \Bl{6}+24 &\Bl{6}+24& (\ref{op:fl2h2d2f}-\ref{op:fl2h2d2l})\\
			\cline{2-7}
			  & (2,2) & \Bl{F^2_{\rm L}F^2_{\rm R}} & \Bl{14} & \Bl{17} &\Bl{17}& (\ref{op:fl2fr2f}-\ref{op:fl2fr2l})\\
			  & & F_{\rm L}F_{\rm R}\psi\psi^{\dagger}{}D & \Bl{27} & \Bl{35}+35 &\Bl{35n^2_f}+35n^2_f& (\ref{op:flfrppgdf}-\ref{op:flfrppgdl})\\
			  & & \psi^2\psi^{\dagger 2}D^2 & \Bl{21} & \Bl{62}+496 & \Bl{\frac12n^2_f(87n^2_f+11)}+n^2_f(348n^2_f+11) & (\ref{op:p2pg2d2f}-\ref{op:p2pg2d2l})\\
			  & & F_{\rm R}\psi^2\phi D^2 & \Bl{8} & \Bl{8}+24 &\Bl{8n^2_f}+24n^2_f& (\ref{op:frp2hd2f}-\ref{op:frp2hd2l})\\
			  & & F_{\rm L} \psi^{\dagger}{}^2 \phi D^2 & \Bl{8} & \Bl{8}+64 &\Bl{8n^2_f}+64n^2_f& (\ref{op:flpg2hd2f}-\ref{op:flpg2hd2l})\\
			  & & F_{\rm L}F_{\rm R}\phi^2D^2 & \Bl{5} & \Bl{6}+12 &\Bl{6}+12& (\ref{op:flfrh2d2f}-\ref{op:flfrh2d2l})\\
			  & & \psi\psi^{\dagger}{}\phi^2D^3 & \Bl{7} & \Bl{16}+135 &\Bl{16n^2_f}+135n_f^2& (\ref{op:ppgh2d3f}-\ref{op:ppgh2d3l})\\
			  & & \phi^4D^4 & \Bl{1} & \Bl{3}+10 &\Bl{3}+10& (\ref{op:h4d4}) \\
			  \cline{2-7}
			  & (1,3) & F_{\rm R} \psi^{\dagger}{}^2 \phi D^2 & \Bl{8} & \Bl{16}+40 &\Bl{16n^2_f}+40n^2_f& (\ref{op:frpg2hd2f}-\ref{op:frpg2hd2l})\\
			  & & F_{\rm R}^2\phi^2 D^2 & \Bl{4} & \Bl{6}+8 &\Bl{6}+8& (\ref{op:fr2h2d2f}-\ref{op:fr2h2d2l})\\
			  & & F^2_{\rm R}\psi\psi^{\dagger}{}D & \Bl{11}+10 & \Bl{11}+59 &\Bl{11n^2_f}+59n^2_f& (\ref{op:fr2ppgdf}-\ref{op:fr2ppgdl})\\
			\hline
			\end{array}\notag}
\end{align}
\caption{Part \uppercase\expandafter{\romannumeral1} of a complete statistics of dimension 8 operators in the Green's basis of SMEFT. $N$ is the number of particles in the class, and $n$, $\tilde{n}$ are defined in eq.~(\ref{eq:ntilden}). $n_f$ denotes the number of generations of fermions. $\mc{N}_{\text{type}},\mc{N}_{\text{term}}$, and $\mc{N}_{\text{operator}}$ give the number of real types, terms and Hermitian operators respectively (independent conjugates are counted). The classes with $N=2,3$ are new classes that only exist in the Green's basis while vanish in the on-shell basis. The classes in blue are classes where no new operator shows up compared to those in the on-shell basis. The numbers in blue of types, terms and operators count the types, terms and operators that are in the on-shell basis, while the numbers in black of types, terms and operators count the new types, terms and operators in the Green's basis.}\label{tab:statistic1}
\end{table}

\begin{table}[htbp]
	\centering	
	\begin{align}{\small
		\begin{array}{cc|c|c|c|c|c}
			\hline
			N & (n,\tilde{n}) & \text{Classes} & \mathcal{N}_{\text{type}} & \mathcal{N}_{\text{term}} & \mathcal{N}_{\text{operator}} & \text{Equations}\\
			\hline\hline
			5 & (3,0) & \Bl{F_{\rm L}\psi^4+h.c.} & \Bl{22} & \Bl{120} &\Bl{2n^3_f(30n_f+1)} & (\ref{op:flp4f}-\ref{op:flp4l})\\
			 & & \Bl{F^2_{\rm L}\psi^2\phi+h.c.} & \Bl{32} & \Bl{60} &\Bl{60n^2_f}& (\ref{op:fl2p2hf}-\ref{op:fl2p2hl})\\
			 & & \Bl{F^3_{\rm L}\phi^2+h.c.} & \Bl{6} & \Bl{6} &\Bl{6}& (\ref{op:fl3h2f}-\ref{op:fl3h2l})\\
			 \cline{2-7}
			 & (2,1) & \Bl{F_{\rm L}\psi^2\psi^{\dagger}{}^2+h.c.} & \Bl{108} & \Bl{204} &\Bl{2n^2_f(71n^2_f-2)}& (\ref{op:flp2pg2f}-\ref{op:flp2pg2l})\\
			 & & \Bl{F^2_{\rm L}\psi^{\dagger 2}\phi+h.c.} & \Bl{32} & \Bl{36} &\Bl{36n^2_f}& (\ref{op:fl2pg2hf}-\ref{op:fl2pg2hl})\\
			 & & \psi^3\psi^{\dagger}{}\phi D+h.c. & \Bl{46} & \Bl{236}+394 & \Bl{2n^3_f(82n_f+1)}+\frac23n^2_f(410n^2_f-3n_f+1) & (\ref{op:p3pghdf}-\ref{op:p3pghdl})\\
			 & & F_{\rm L}\psi\psi^{\dagger}{}\phi^2D & \Bl{19} & \Bl{46}+44 &\Bl{46n^2_f}+44n^2_f& (\ref{op:flppgh2df}-\ref{op:flppgh2dl})\\
			 & & \psi^2\phi^3D^2+h.c. & \Bl{6} & \Bl{36}+60 &\Bl{36n^2_f}+60n^2_f& (\ref{op:p2h3d2f}-\ref{op:p2h3d2l})\\
			 & & F_{\rm L}\phi^4D^2 & \Bl{2} & \Bl{3}+4 &\Bl{3}+4& (\ref{op:flh4d2f}-\ref{op:flh4d2l})\\
			 \cline{2-7}
			 & (1,2) & F_{\rm R}\psi\psi^{\dagger}{}\phi^2D & \Bl{19} & \Bl{46}+22 &\Bl{46n^2_f}+22n^2_f& (\ref{op:frppgh2df}-\ref{op:frppgh2dl})\\
			 & & \Bl{F_{\rm R}\phi^4D^2} & \Bl{2} & \Bl{3} &\Bl{3}& (\ref{op:frh4d2f}-\ref{op:frh4d2l})\\
			\hline
			6 & (2,0) & \Bl{\psi^4\phi^2+h.c.} & \Bl{16} & \Bl{66} & \Bl{\frac13 n_f^2 (91 n_f^2+17)}& (\ref{op:p4h2f}-\ref{op:p4h2l})\\
			 & & \Bl{F_{\rm L}\psi^2\phi^3+h.c.} & \Bl{16} & \Bl{22} &\Bl{22n^2_f}& (\ref{op:flp2h3f}-\ref{op:flp2h3l})\\
			 & & \Bl{F^2_{\rm L}\phi^4+h.c.} & \Bl{8} & \Bl{10} &\Bl{10}& (\ref{op:fl2h4f}-\ref{op:fl2h4l})\\
			 \cline{2-7}
			 & (1,1) & \Bl{\psi^2\psi^{\dagger}{}^2\phi^2} & \Bl{33} & \Bl{71} & \Bl{n^2_f(51n^2_f-2n_f+2)} & (\ref{op:p2pg2h2f}-\ref{op:p2pg2h2l})\\
			 & & \psi\psi^{\dagger}{}\phi^4D & \Bl{7} & \Bl{13}+14 & \Bl{13n_f^2}+14n_f^2 & (\ref{op:ppgh4df}-\ref{op:ppgh4dl})\\
			 & & \phi^6D^2 & \Bl{1} & \Bl{2}+2 &\Bl{2}+2& (\ref{op:h6d2})\\
			\hline
			7 & (1,0) & \Bl{\psi^2\phi^5+h.c.} & \Bl{6} & \Bl{6} & \Bl{6n_f^2} & (\ref{op:p2h5f}-\ref{op:p2h5l})\\
			\hline
			8 & (0,0) & \Bl{\phi^8} & \Bl{1} & \Bl{1} &\Bl{1}& (\ref{op:h8})\\
			\hline\hline
			 \multicolumn{2}{c|}{\multirow{2}*{\text{Total}}} & \multirow{2}*{59}  & \multirow{2}*{\Bl{541}+69} & \multirow{2}*{\Bl{1266}+1998} & \Bl{\frac{1}{6}(3047 n_f^4+6n_f^3+2371 n_f^2+534)} \\ & & & & & +\frac{1}{3}(2170 n_f^4-6n_f^3+2525 n_f^2+258) \\
			 \hline
	\end{array}\notag}
\end{align}
\caption{Part \uppercase\expandafter{\romannumeral2} of a complete statistics of dimension 8 operators in the Green's basis of SMEFT. $N$ is the number of particles in the class, and $n$, $\tilde{n}$ are defined in eq.~(\ref{eq:ntilden}). $n_f$ denotes the number of generations of fermions. $\mc{N}_{\text{type}},\mc{N}_{\text{term}}$, and $\mc{N}_{\text{operator}}$ give the number of real types, terms and Hermitian operators respectively (independent conjugates are counted). The classes in blue are classes where no new operator shows up compared to those in the on-shell basis. The numbers in blue of types, terms and operators count the types, terms and operators that are in the on-shell basis, while the numbers in black of types, terms and operators count the new types, terms and operators in the Green's basis.}\label{tab:statistic2}
\end{table}

\subsection{Pure Bosonic Operators}\label{sec:bosonopes}

The bosonic operators in the Green's basis in the SMEFT have been listed in Ref.~\cite{Chala:2021cgt}. In this subsection, we list an alternative basis of the bosonic operators which is equivalent with the basis in Ref.~\cite{Chala:2021cgt}. It should be noted that the $F_{\rm L}/F_{\rm R}$ basis is used in our operator basis, so the operators is connected with those in the reference by eq.~(\ref{eq:FLR1}). For example, there are 12 independent operators in $B^2 \phi^2 D^2$ in Ref.~\cite{Chala:2021cgt}, which correspond to the 6 operators in type $B^2_{\rm{L}} H H^{\dagger}{} D^2$, the 3 operators in type $B_{\rm{L}} H H^{\dagger}{}  B_{\rm{R}}D^2$ and the 3 operators in type $H H^{\dagger}{}  B^2_{\rm{R}}D^2$.\\

\underline{Class $\Bl{F_{\rm{L}}^4 }$}: 7 types
\begin{align}\label{op:fl4f}
	
	&\begin{array}{c|l}
		
		\multirow{1}*{$\Bl{\mathcal{O}_{B_{\rm{L}}^4}}$}
		
		&\Bl{ B_{\rm{L}}{}{}_{\lambda \rho }{}B_{\rm{L}}{}{}_{\mu \nu }{}B_{\rm{L}}{}{}{}^{\lambda \rho }B_{\rm{L}}{}{}{}^{\mu \nu }}
		
	\end{array}\\
	
	&\begin{array}{c|l}
		
		\multirow{1}*{$\Bl{\mathcal{O}_{B_{\rm{L}}^2 W_{\rm{L}}^2}^{\left(1\sim2\right)}}$}
		
		&\Bl{ B_{\rm{L}}{}{}_{\mu \nu }{}B_{\rm{L}}{}{}{}^{\mu \nu }W_{\rm{L}}^I{}{}_{\lambda \rho }{}W_{\rm{L}}^I{}{}^{\lambda \rho }},
		
		\quad\Bl{ B_{\rm{L}}{}{}_{\lambda \rho }{}B_{\rm{L}}{}{}_{\mu \nu }{}W_{\rm{L}}^I{}{}^{\lambda \rho }W_{\rm{L}}^I{}{}^{\mu \nu }}
		
	\end{array}\\
	
	&\begin{array}{c|l}
		
		\multirow{1}*{$\Bl{\mathcal{O}_{W_{\rm{L}}^4}^{\left(1\sim2\right)}}$}
		
		&\Bl{ W_{\rm{L}}^I{}_{\lambda \rho }{}W_{\rm{L}}^I{}_{\mu \nu }{}W_{\rm{L}}^J{}{}^{\lambda \rho }W_{\rm{L}}^J{}{}^{\mu \nu }},
		
		\quad\Bl{ W_{\rm{L}}^I{}_{\mu \nu }{}W_{\rm{L}}^I{}{}^{\mu \nu }W_{\rm{L}}^J{}{}_{\lambda \rho }{}W_{\rm{L}}^J{}{}^{\lambda \rho }}
		
	\end{array}\\
	
	&\begin{array}{c|l}
		
		\multirow{1}*{$\Bl{\mathcal{O}_{B_{\rm{L}}^2 G_{\rm{L}}^2}^{\left(1\sim2\right)}}$}
		
		&\Bl{ B_{\rm{L}}{}{}_{\mu \nu }{}B_{\rm{L}}{}{}{}^{\mu \nu }G_{\rm{L}}^A{}{}_{\lambda \rho }{}G_{\rm{L}}^A{}{}^{\lambda \rho }},
		
		\quad\Bl{ B_{\rm{L}}{}{}_{\lambda \rho }{}B_{\rm{L}}{}{}_{\mu \nu }{}G_{\rm{L}}^A{}{}^{\lambda \rho }G_{\rm{L}}^A{}{}^{\mu \nu }}
		
	\end{array}\\
	
	&\begin{array}{c|l}
		
		\multirow{1}*{$\Bl{\mathcal{O}_{G_{\rm{L}}^2 W_{\rm{L}}^2}^{\left(1\sim2\right)}}$}
		
		&\Bl{ G_{\rm{L}}^A{}_{\mu \nu }{}G_{\rm{L}}^A{}{}^{\mu \nu }W_{\rm{L}}^I{}{}_{\lambda \rho }{}W_{\rm{L}}^I{}{}^{\lambda \rho }},
		
		\quad\Bl{ G_{\rm{L}}^A{}_{\lambda \rho }{}G_{\rm{L}}^A{}_{\mu \nu }{}W_{\rm{L}}^I{}{}^{\lambda \rho }W_{\rm{L}}^I{}{}^{\mu \nu }}
		
	\end{array}\\
	
	&\begin{array}{c|l}
		
		\multirow{1}*{$\Bl{\mathcal{O}_{B_{\rm{L}} G_{\rm{L}}^3}}$}
		
		&\Bl{ d{}^{ABC}B_{\rm{L}}{}{}{}_{\mu \nu }{}G_{\rm{L}}^A{}{}^{\mu \nu }G_{\rm{L}}^B{}{}_{\lambda \rho }{}G_{\rm{L}}^C{}{}^{\lambda \rho }}
		
	\end{array}\\
	
	&\begin{array}{c|l}
		
		\multirow{2}*{$\Bl{\mathcal{O}_{G_{\rm{L}}^4}^{\left(1\sim3\right)}}$}
		
		&\Bl{ d{}^{ABE}d{}^{CDE}G_{\rm{L}}^A{}{}_{\mu \nu }{}G_{\rm{L}}^B{}{}^{\mu \nu }G_{\rm{L}}^C{}{}_{\lambda \rho }{}G_{\rm{L}}^D{}{}^{\lambda \rho }},
		
		\quad\Bl{ d{}^{ABE}d{}^{CDE}G_{\rm{L}}^A{}{}_{\mu \nu }{}G_{\rm{L}}^B{}_{\lambda \rho }{}G_{\rm{L}}^C{}{}^{\mu \nu }G_{\rm{L}}^D{}{}^{\lambda \rho }},
		
		\\&\Bl{ f{}^{ABE}f{}^{CDE}G_{\rm{L}}^A{}{}_{\mu \nu }{}G_{\rm{L}}^B{}_{\lambda \rho }{}G_{\rm{L}}^C{}{}^{\mu \nu }G_{\rm{L}}^D{}{}^{\lambda \rho }}
		
	\end{array}\label{op:fl4l}
	
\end{align}

\underline{Class $\Bl{F_{\rm{L}}^2 F_{\rm{R}}^2 }$}: 10 types
\begin{align}\label{op:fl2fr2f}
	
	&\begin{array}{c|l}
		
		\multirow{1}*{$\Bl{\mathcal{O}_{B_{\rm{L}}^2 B_{\rm{R}}^2}}$}
		
		&\Bl{ B_{\rm{L}}{}{}_{\mu \nu }{}B_{\rm{L}}{}{}{}^{\mu \nu }B_{\rm{R}}{}{}{}_{\lambda \rho }{}B_{\rm{R}}{}{}{}^{\lambda \rho }}
		
	\end{array}\\
	
	&\begin{array}{c|l}
		
		\multirow{1}*{$\Bl{\mathcal{O}_{B_{\rm{L}}^2 W_{\rm{R}}^2}}$}
		
		&\Bl{ B_{\rm{L}}{}{}_{\mu \nu }{}B_{\rm{L}}{}{}{}^{\mu \nu }W_{\rm{R}}^I{}{}_{\lambda \rho }{}W_{\rm{R}}^I{}{}^{\lambda \rho }}
		
	\end{array}\\
	
	&\begin{array}{c|l}
		
		\multirow{1}*{$\Bl{\mathcal{O}_{B_{\rm{L}} W_{\rm{L}} B_{\rm{R}} W_{\rm{R}}}}$}
		
		&\Bl{ B_{\rm{L}}{}{}_{\mu \nu }{}B_{\rm{R}}{}{}_{\lambda \rho }{}W_{\rm{L}}^I{}{}^{\mu \nu }W_{\rm{R}}^I{}{}^{\lambda \rho }}
		
	\end{array}\\
	
	&\begin{array}{c|l}
		
		\multirow{1}*{$\Bl{\mathcal{O}_{B_{\rm{L}}^2 G_{\rm{R}}^2}}$}
		
		&\Bl{ B_{\rm{L}}{}{}_{\mu \nu }{}B_{\rm{L}}{}{}{}^{\mu \nu }G_{\rm{R}}^A{}{}_{\lambda \rho }{}G_{\rm{R}}^A{}{}^{\lambda \rho }}
		
	\end{array}\\
	
	&\begin{array}{c|l}
		
		\multirow{1}*{$\Bl{\mathcal{O}_{B_{\rm{L}} G_{\rm{L}} B_{\rm{R}} G_{\rm{R}}}}$}
		
		&\Bl{ B_{\rm{L}}{}{}_{\mu \nu }{}B_{\rm{R}}{}{}_{\lambda \rho }{}G_{\rm{L}}^A{}{}^{\mu \nu }G_{\rm{R}}^A{}{}^{\lambda \rho }}
		
	\end{array}\\
	
	&\begin{array}{c|l}
		
		\multirow{1}*{$\Bl{\mathcal{O}_{W_{\rm{L}}^2 W_{\rm{R}}^2}^{\left(1\sim2\right)}}$}
		
		&\Bl{ W_{\rm{L}}^I{}_{\mu \nu }{}W_{\rm{L}}^J{}{}^{\mu \nu }W_{\rm{R}}^I{}{}_{\lambda \rho }{}W_{\rm{R}}^J{}{}^{\lambda \rho }},
		
		\quad\Bl{ W_{\rm{L}}^I{}_{\mu \nu }{}W_{\rm{L}}^I{}{}^{\mu \nu }W_{\rm{R}}^K{}{}_{\lambda \rho }{}W_{\rm{R}}^K{}{}^{\lambda \rho }}
		
	\end{array}\\
	
	&\begin{array}{c|l}
		
		\multirow{1}*{$\Bl{\mathcal{O}_{W_{\rm{L}}^2 G_{\rm{R}}^2}}$}
		
		&\Bl{ G_{\rm{R}}^A{}_{\lambda \rho }{}G_{\rm{R}}^A{}{}^{\lambda \rho }W_{\rm{L}}^I{}{}_{\mu \nu }{}W_{\rm{L}}^I{}{}^{\mu \nu }}
		
	\end{array}\\
	
	&\begin{array}{c|l}
		
		\multirow{1}*{$\Bl{\mathcal{O}_{G_{\rm{L}} W_{\rm{L}} G_{\rm{R}} W_{\rm{R}}}}$}
		
		&\Bl{ G_{\rm{L}}^A{}_{\mu \nu }{}G_{\rm{R}}^A{}_{\lambda \rho }{}W_{\rm{L}}^I{}{}^{\mu \nu }W_{\rm{R}}^I{}{}^{\lambda \rho }}
		
	\end{array}\\
	
	&\begin{array}{c|l}
		
		\multirow{1}*{$\Bl{\mathcal{O}_{B_{\rm{L}} G_{\rm{L}} G_{\rm{R}}^2}}$}
		
		&\Bl{ d{}^{ABC}B_{\rm{L}}{}{}{}_{\mu \nu }{}G_{\rm{L}}^A{}{}^{\mu \nu }G_{\rm{R}}^B{}{}_{\lambda \rho }{}G_{\rm{R}}^C{}{}^{\lambda \rho }}
		
	\end{array}\\
	
	&\begin{array}{c|l}
		
		\multirow{2}*{$\Bl{\mathcal{O}_{G_{\rm{L}}^2 G_{\rm{R}}^2}^{\left(1\sim3\right)}}$}
		
		&\Bl{ d{}^{ABE}d{}^{CDE}G_{\rm{L}}^A{}{}_{\mu \nu }{}G_{\rm{L}}^B{}{}^{\mu \nu }G_{\rm{R}}^C{}{}_{\lambda \rho }{}G_{\rm{R}}^D{}{}^{\lambda \rho }},
		
		\quad\Bl{ G_{\rm{L}}^A{}_{\mu \nu }{}G_{\rm{L}}^A{}{}^{\mu \nu }G_{\rm{R}}^C{}{}_{\lambda \rho }{}G_{\rm{R}}^C{}{}^{\lambda \rho }},
		
		\\&\Bl{ G_{\rm{L}}^A{}_{\mu \nu }{}G_{\rm{L}}^B{}{}^{\mu \nu }G_{\rm{R}}^A{}{}_{\lambda \rho }{}G_{\rm{R}}^B{}{}^{\lambda \rho }}
		
	\end{array}\label{op:fl2fr2l}
	
\end{align}

\underline{Class $F_{\rm{L}}^3 D^2$}: 4 types
\begin{align}\label{op:fl3d2f}
	
	&\begin{array}{c|l}
		
		\multirow{1}*{$\mathcal{O}_{B_{\rm{L}} W_{\rm{L}}^2D^2}^{\left(1\sim2\right)}$}
		
		& W_{\rm{L}}^I{}{}^{\lambda \rho }B_{\rm{L}}{}{}{}_{\lambda }{}{}^{\nu } D_{\nu } D_{\mu } W_{\rm{L}}^I{}{}_{\rho }{}{}^{\mu } ,
		
		\quad B_{\rm{L}}{}{}_{\lambda }{}{}^{\nu } D_{\mu } W_{\rm{L}}^I{}{}_{\rho }{}{}^{\mu }  D_{\nu } W_{\rm{L}}^I{}{}^{\lambda \rho } 
		
	\end{array}\\
	
	&\begin{array}{c|l}
		
		\multirow{1}*{$\mathcal{O}_{W_{\rm{L}}^3D^2}^{\left(1\sim2\right)}$}
		
		& \epsilon {}^{IJK}W_{\rm{L}}^K{}{}^{\lambda \rho }W_{\rm{L}}^I{}{}_{\lambda }{}{}^{\nu } D_{\nu } D_{\mu } W_{\rm{L}}^J{}{}_{\rho }{}{}^{\mu } ,
		
		\quad \epsilon {}^{IJK}W_{\rm{L}}^I{}{}^{\nu \mu } D_{\mu } W_{\rm{L}}^J{}{}_{\lambda \rho }{}  D_{\nu } W_{\rm{L}}^K{}{}^{\lambda \rho } 
		
	\end{array}\\
	
	&\begin{array}{c|l}
		
		\multirow{1}*{$\mathcal{O}_{B_{\rm{L}} G_{\rm{L}}^2D^2}^{\left(1\sim2\right)}$}
		
		& G_{\rm{L}}^A{}{}^{\lambda \rho }B_{\rm{L}}{}{}{}_{\lambda }{}{}^{\nu } D_{\nu } D_{\mu } G_{\rm{L}}^A{}{}_{\rho }{}{}^{\mu } ,
		
		\quad B_{\rm{L}}{}{}_{\lambda }{}{}^{\nu } D_{\nu } G_{\rm{L}}^A{}{}^{\lambda \rho }  D_{\mu } G_{\rm{L}}^A{}{}_{\rho }{}{}^{\mu } 
		
	\end{array}\\
	
	&\begin{array}{c|l}
		
		\multirow{1}*{$\mathcal{O}_{G_{\rm{L}}^3D^2}^{\left(1\sim2\right)}$}
		
		& f{}^{ABC}G_{\rm{L}}^C{}{}^{\lambda \rho }G_{\rm{L}}^A{}{}_{\lambda }{}{}^{\nu } D_{\nu } D_{\mu } G_{\rm{L}}^B{}{}_{\rho }{}{}^{\mu } ,
		
		\quad f{}^{ABC}G_{\rm{L}}^A{}{}^{\nu \mu } D_{\mu } G_{\rm{L}}^B{}{}_{\lambda \rho }{}  D_{\nu } G_{\rm{L}}^C{}{}^{\lambda \rho } 
		
	\end{array}\label{op:fl3d2l}
	
\end{align}

\underline{Class $F_{\rm{L}}^2 F_{\rm{R}} D^2$}: 4 types
\begin{align}\label{op:fl2frd2f}
	
	&\begin{array}{c|l}
		
		\multirow{1}*{$\mathcal{O}_{B_{\rm{L}} W_{\rm{L}} W_{\rm{R}}D^2}$}
		
		& W_{\rm{R}}^I{}{}^{\lambda \rho }B_{\rm{L}}{}{}{}_{\lambda }{}{}^{\nu } D_{\nu } D_{\mu } W_{\rm{L}}^I{}{}_{\rho }{}{}^{\mu } 
		
	\end{array}\\
	
	&\begin{array}{c|l}
		
		\multirow{1}*{$\mathcal{O}_{W_{\rm{L}}^2 W_{\rm{R}}D^2}$}
		
		& \epsilon {}^{IJK}W_{\rm{R}}^K{}{}^{\lambda \rho }W_{\rm{L}}^I{}{}_{\lambda }{}{}^{\nu } D_{\nu } D_{\mu } W_{\rm{L}}^J{}{}_{\rho }{}{}^{\mu } 
		
	\end{array}\\
	
	&\begin{array}{c|l}
		
		\multirow{1}*{$\mathcal{O}_{B_{\rm{L}} G_{\rm{L}} G_{\rm{R}}D^2}$}
		
		& G_{\rm{R}}^A{}{}^{\lambda \rho }B_{\rm{L}}{}{}{}_{\lambda }{}{}^{\nu } D_{\nu } D_{\mu } G_{\rm{L}}^A{}{}_{\rho }{}{}^{\mu } 
		
	\end{array}\\
	
	&\begin{array}{c|l}
		
		\multirow{1}*{$\mathcal{O}_{G_{\rm{L}}^2 G_{\rm{R}}D^2}$}
		
		& f{}^{ABC}G_{\rm{R}}^C{}{}^{\lambda \rho }G_{\rm{L}}^A{}{}_{\lambda }{}{}^{\nu } D_{\nu } D_{\mu } G_{\rm{L}}^B{}{}_{\rho }{}{}^{\mu } 
		
	\end{array}\label{op:fl2frd2l}
	
\end{align}

\underline{Class $F_{\rm{L}} F_{\rm{R}}^2 D^2$}: 4 types
\begin{align}\label{op:flfr2d2f}
	
	&\begin{array}{c|l}
		
		\multirow{1}*{$\mathcal{O}_{W_{\rm{L}} B_{\rm{R}} W_{\rm{R}}D^2}$}
		
		& W_{\rm{L}}^I{}_{\lambda }{}{}^{\mu } D_{\mu } B_{\rm{R}}{}{}{}_{\rho }{}{}^{\nu }  D_{\nu } W_{\rm{R}}^I{}{}^{\lambda \rho } 
		
	\end{array}\\
	
	&\begin{array}{c|l}
		
		\multirow{1}*{$\mathcal{O}_{G_{\rm{L}} B_{\rm{R}} G_{\rm{R}}D^2}$}
		
		& G_{\rm{L}}^A{}_{\lambda }{}{}^{\mu } D_{\nu } G_{\rm{R}}^A{}{}^{\lambda \rho }  D_{\mu } B_{\rm{R}}{}{}{}_{\rho }{}{}^{\nu } 
		
	\end{array}\\
	
	&\begin{array}{c|l}
		
		\multirow{1}*{$\mathcal{O}_{W_{\rm{L}} W_{\rm{R}}^2D^2}$}
		
		& \epsilon {}^{IJK}W_{\rm{L}}^I{}{}_{\lambda }{}{}^{\mu } D_{\mu } W_{\rm{R}}^J{}{}_{\rho }{}{}^{\nu }  D_{\nu } W_{\rm{R}}^K{}{}^{\lambda \rho } 
		
	\end{array}\\
	
	&\begin{array}{c|l}
		
		\multirow{1}*{$\mathcal{O}_{G_{\rm{L}} G_{\rm{R}}^2D^2}$}
		
		& f{}^{ABC}G_{\rm{L}}^A{}{}_{\lambda }{}{}^{\mu } D_{\nu } G_{\rm{R}}^C{}{}^{\lambda \rho }  D_{\mu } G_{\rm{R}}^B{}{}_{\rho }{}{}^{\nu } 
		
	\end{array}\label{op:flfr2d2l}
	
\end{align}

\underline{Class $\Bl{F_{\rm{L}}^3 \phi ^2 }$}: 3 types
\begin{align}\label{op:fl3h2f}
	
	&\begin{array}{c|l}
		
		\multirow{1}*{$\Bl{\mathcal{O}_{B_{\rm{L}} W_{\rm{L}}^2 H H^{\dagger}{} }}$}
		
		&\Bl{ (\tau {}^K)_j^i\epsilon {}^{IJK}H_i{}H^{\dagger} {}^jB_{\rm{L}}{}{}{}_{\mu \nu }{}W_{\rm{L}}^J{}{}^{\nu \lambda }W_{\rm{L}}^I{}{}^{\mu }{}_{\lambda }{}}
		
	\end{array}\\
	
	&\begin{array}{c|l}
		
		\multirow{1}*{$\Bl{\mathcal{O}_{W_{\rm{L}}^3 H H^{\dagger}{} }}$}
		
		&\Bl{  \epsilon {}^{IJK}H_i{}H^{\dagger} {}^iW_{\rm{L}}^I{}{}_{\mu \nu }{}W_{\rm{L}}^K{}{}^{\nu \lambda }W_{\rm{L}}^J{}{}^{\mu }{}_{\lambda }{}}
		
	\end{array}\\
	
	&\begin{array}{c|l}
		
		\multirow{1}*{$\Bl{\mathcal{O}_{G_{\rm{L}}^3 H H^{\dagger}{} }}$}
		
		&\Bl{ f{}^{ABC}H_i{}H^{\dagger} {}^iG_{\rm{L}}^A{}{}_{\mu \nu }{}G_{\rm{L}}^C{}{}^{\nu \lambda }G_{\rm{L}}^B{}{}^{\mu }{}_{\lambda }{}}
		
	\end{array}\label{op:fl3h2l}
	
\end{align}

\underline{Class $F_{\rm{L}}^2 D^4$}: 3 types
\begin{align}\label{op:fl2d4f}
	
	&\begin{array}{c|l}
		
		\multirow{1}*{$\mathcal{O}_{B_{\rm{L}}^2D^4}$}
		
		& B_{\rm{L}}{}{}_{\lambda \rho }{} D_{\mu } D^{\mu } D^{\nu } D_{\nu } B_{\rm{L}}{}{}{}^{\lambda \rho } 
		
	\end{array}\\
	
	&\begin{array}{c|l}
		
		\multirow{1}*{$\mathcal{O}_{W_{\rm{L}}^2D^4}$}
		
		& W_{\rm{L}}^I{}_{\lambda \rho }{} D_{\mu } D^{\mu } D^{\nu } D_{\nu } W_{\rm{L}}^I{}{}^{\lambda \rho } 
		
	\end{array}\\
	
	&\begin{array}{c|l}
		
		\multirow{1}*{$\mathcal{O}_{G_{\rm{L}}^2D^4}$}
		
		& G_{\rm{L}}^A{}_{\lambda \rho }{} D_{\mu } D^{\mu } D^{\nu } D_{\nu } G_{\rm{L}}^A{}{}^{\lambda \rho } 
		
	\end{array}\label{op:fl2d4l}
	
\end{align}

\underline{Class $F_{\rm{L}}^2 \phi ^2 D^2$}: 4 types
\begin{align}\label{op:fl2h2d2f}
	
	&\begin{array}{c|l}
		
		\multirow{3}*{$\mathcal{O}_{B_{\rm{L}}^2 H H^{\dagger}{} D^2}^{\left(1\sim6\right)}$}
		
		&\Bl{ B_{\rm{L}}{}{}{}^{\lambda \mu }B_{\rm{L}}{}{}{}_{\lambda }{}{}^{\nu } D_{\mu } H_i{}  D_{\nu } H^{\dagger} {}^i },
		
		\quad H_i{}H^{\dagger} {}^iB_{\rm{L}}{}{}{}_{\nu \lambda }{} D^{\mu } D_{\mu } B_{\rm{L}}{}{}{}^{\nu \lambda } ,
		
		\\& H^{\dagger} {}^iB_{\rm{L}}{}{}{}_{\nu \lambda }{}B_{\rm{L}}{}{}{}^{\nu \lambda } D^{\mu } D_{\mu } H_i{} ,
		
		\quad H_i{}B_{\rm{L}}{}{}_{\nu \lambda }{}B_{\rm{L}}{}{}{}^{\nu \lambda } D^{\mu } D_{\mu } H^{\dagger} {}^i ,
		
		\\& H^{\dagger} {}^iB_{\rm{L}}{}{}{}_{\lambda }{}{}^{\nu } D_{\nu } H_i{}  D_{\mu } B_{\rm{L}}{}{}{}^{\lambda \mu } ,
		
		\quad H_i{}B_{\rm{L}}{}{}_{\lambda }{}{}^{\nu } D_{\nu } H^{\dagger} {}^i  D_{\mu } B_{\rm{L}}{}{}{}^{\lambda \mu } 
		
	\end{array}\\
	
	&\begin{array}{c|l}
		
		\multirow{5}*{$\mathcal{O}_{B_{\rm{L}} W_{\rm{L}} H H^{\dagger}{} D^2}^{\left(1\sim9\right)}$}
		
		&\Bl{ (\tau {}^I)_j^iW_{\rm{L}}^I{}{}^{\lambda \mu }B_{\rm{L}}{}{}{}_{\lambda }{}{}^{\nu } D_{\mu } H_i{}  D_{\nu } H^{\dagger} {}^j },
		
		\quad\Bl{ (\tau {}^I)_j^iH_i{}B_{\rm{L}}{}{}_{\nu \lambda }{} D^{\mu } H^{\dagger} {}^j  D_{\mu } W_{\rm{L}}^I{}{}^{\nu \lambda } },
		
		\\& (\tau {}^I)_j^iH_i{}H^{\dagger} {}^jB_{\rm{L}}{}{}{}_{\nu \lambda }{} D^{\mu } D_{\mu } W_{\rm{L}}^I{}{}^{\nu \lambda } ,
		
		\quad (\tau {}^I)_j^iH^{\dagger} {}^jB_{\rm{L}}{}{}{}_{\nu \lambda }{}W_{\rm{L}}^I{}{}^{\nu \lambda } D^{\mu } D_{\mu } H_i{} ,
		
		\\& (\tau {}^I)_j^iH_i{}B_{\rm{L}}{}{}_{\nu \lambda }{}W_{\rm{L}}^I{}{}^{\nu \lambda } D^{\mu } D_{\mu } H^{\dagger} {}^j ,
		
		\quad (\tau {}^I)_j^iH^{\dagger} {}^jB_{\rm{L}}{}{}{}_{\nu \lambda }{} D^{\mu } H_i{}  D_{\mu } W_{\rm{L}}^I{}{}^{\nu \lambda } ,
		
		\\& (\tau {}^I)_j^iH^{\dagger} {}^jB_{\rm{L}}{}{}{}_{\lambda }{}{}^{\nu } D_{\nu } H_i{}  D_{\mu } W_{\rm{L}}^I{}{}^{\lambda \mu } ,
		
		\quad (\tau {}^I)_j^iH_i{}B_{\rm{L}}{}{}_{\lambda }{}{}^{\nu } D_{\nu } H^{\dagger} {}^j  D_{\mu } W_{\rm{L}}^I{}{}^{\lambda \mu } ,
		
		\\& (\tau {}^I)_j^iB_{\rm{L}}{}{}_{\nu \lambda }{}W_{\rm{L}}^I{}{}^{\nu \lambda } D_{\mu } H_i{}  D^{\mu } H^{\dagger} {}^j 
		
	\end{array}\\
	
	&\begin{array}{c|l}
		
		\multirow{5}*{$\mathcal{O}_{W_{\rm{L}}^2 H H^{\dagger}{} D^2}^{\left(1\sim9\right)}$}
		
		&\Bl{ W_{\rm{L}}^I{}{}^{\lambda \mu }W_{\rm{L}}^I{}{}_{\lambda }{}{}^{\nu } D_{\mu } H_i{}  D_{\nu } H^{\dagger} {}^i },
		
		\quad\Bl{ (\tau {}^K)_j^i\epsilon {}^{IJK}W_{\rm{L}}^J{}{}^{\lambda \mu }W_{\rm{L}}^I{}{}_{\lambda }{}{}^{\nu } D_{\mu } H_i{}  D_{\nu } H^{\dagger} {}^j },
		
		\\& H_i{}H^{\dagger} {}^iW_{\rm{L}}^I{}{}_{\nu \lambda }{} D^{\mu } D_{\mu } W_{\rm{L}}^I{}{}^{\nu \lambda } ,
		
		\quad H^{\dagger} {}^iW_{\rm{L}}^I{}{}_{\nu \lambda }{}W_{\rm{L}}^I{}{}^{\nu \lambda } D^{\mu } D_{\mu } H_i{} ,
		
		\\& H_i{}W_{\rm{L}}^I{}_{\nu \lambda }{}W_{\rm{L}}^I{}{}^{\nu \lambda } D^{\mu } D_{\mu } H^{\dagger} {}^i ,
		
		\quad H^{\dagger} {}^iW_{\rm{L}}^I{}{}_{\lambda }{}{}^{\nu } D_{\nu } H_i{}  D_{\mu } W_{\rm{L}}^I{}{}^{\lambda \mu } ,
		
		\\& H_i{}W_{\rm{L}}^I{}_{\lambda }{}{}^{\nu } D_{\nu } H^{\dagger} {}^i  D_{\mu } W_{\rm{L}}^I{}{}^{\lambda \mu } ,
		
		\quad (\tau {}^K)_j^i\epsilon {}^{IJK}H_i{}H^{\dagger} {}^jW_{\rm{L}}^I{}{}_{\nu \lambda }{} D^{\mu } D_{\mu } W_{\rm{L}}^J{}{}^{\nu \lambda } ,
		
		\\& (\tau {}^K)_j^i\epsilon {}^{IJK}H^{\dagger} {}^jW_{\rm{L}}^I{}{}_{\nu \lambda }{} D^{\mu } H_i{}  D_{\mu } W_{\rm{L}}^J{}{}^{\nu \lambda } 
		
	\end{array}\\
	
	&\begin{array}{c|l}
		
		\multirow{3}*{$\mathcal{O}_{G_{\rm{L}}^2 H H^{\dagger}{} D^2}^{\left(1\sim6\right)}$}
		
		&\Bl{ G_{\rm{L}}^A{}{}^{\lambda \mu }G_{\rm{L}}^A{}{}_{\lambda }{}{}^{\nu } D_{\mu } H_i{}  D_{\nu } H^{\dagger} {}^i },
		
		\quad H_i{}H^{\dagger} {}^iG_{\rm{L}}^A{}{}_{\nu \lambda }{} D^{\mu } D_{\mu } G_{\rm{L}}^A{}{}^{\nu \lambda } ,
		
		\\& H^{\dagger} {}^iG_{\rm{L}}^A{}{}_{\nu \lambda }{}G_{\rm{L}}^A{}{}^{\nu \lambda } D^{\mu } D_{\mu } H_i{} ,
		
		\quad H_i{}G_{\rm{L}}^A{}_{\nu \lambda }{}G_{\rm{L}}^A{}{}^{\nu \lambda } D^{\mu } D_{\mu } H^{\dagger} {}^i ,
		
		\\& H^{\dagger} {}^iG_{\rm{L}}^A{}{}_{\lambda }{}{}^{\nu } D_{\nu } H_i{}  D_{\mu } G_{\rm{L}}^A{}{}^{\lambda \mu } ,
		
		\quad H_i{}G_{\rm{L}}^A{}_{\lambda }{}{}^{\nu } D_{\nu } H^{\dagger} {}^i  D_{\mu } G_{\rm{L}}^A{}{}^{\lambda \mu } 
		
	\end{array}\label{op:fl2h2d2l}
	
\end{align}

\underline{Class $F_{\rm{L}} F_{\rm{R}} \phi ^2 D^2$}: 4 types
\begin{align}\label{op:flfrh2d2f}
	
	&\begin{array}{c|l}
		
		\multirow{2}*{$\mathcal{O}_{B_{\rm{L}} H H^{\dagger}{}  B_{\rm{R}}D^2}^{\left(1\sim3\right)}$}
		
		&\Bl{ H_i{}B_{\rm{R}}{}{}{}^{\lambda \mu }B_{\rm{L}}{}{}{}_{\lambda }{}{}^{\nu } D_{\nu } D_{\mu } H^{\dagger} {}^i },
		
		\quad H^{\dagger} {}^iB_{\rm{R}}{}{}{}^{\lambda \mu }B_{\rm{L}}{}{}{}_{\lambda }{}{}^{\nu } D_{\nu } D_{\mu } H_i{} ,
		
		\\& B_{\rm{R}}{}{}{}^{\lambda \mu }B_{\rm{L}}{}{}{}_{\lambda }{}{}^{\nu } D_{\mu } H_i{}  D_{\nu } H^{\dagger} {}^i 
		
	\end{array}\\
	
	&\begin{array}{c|l}
		
		\multirow{2}*{$\mathcal{O}_{B_{\rm{L}} H H^{\dagger}{}  W_{\rm{R}}D^2}^{\left(1\sim3\right)}$}
		
		&\Bl{ (\tau {}^I)_j^iH_i{}W_{\rm{R}}^I{}{}^{\lambda \mu }B_{\rm{L}}{}{}{}_{\lambda }{}{}^{\nu } D_{\nu } D_{\mu } H^{\dagger} {}^j },
		
		\quad (\tau {}^I)_j^iH^{\dagger} {}^jW_{\rm{R}}^I{}{}^{\lambda \mu }B_{\rm{L}}{}{}{}_{\lambda }{}{}^{\nu } D_{\nu } D_{\mu } H_i{} ,
		
		\\& (\tau {}^I)_j^iW_{\rm{R}}^I{}{}^{\lambda \mu }B_{\rm{L}}{}{}{}_{\lambda }{}{}^{\nu } D_{\mu } H_i{}  D_{\nu } H^{\dagger} {}^j 
		
	\end{array}\\
	
	&\begin{array}{c|l}
		
		\multirow{3}*{$\mathcal{O}_{W_{\rm{L}} H H^{\dagger}{}  W_{\rm{R}}D^2}^{\left(1\sim6\right)}$}
		
		&\Bl{ (\tau {}^K)_j^i\epsilon {}^{IJK}H_i{}W_{\rm{R}}^J{}{}^{\lambda \mu }W_{\rm{L}}^I{}{}_{\lambda }{}{}^{\nu } D_{\nu } D_{\mu } H^{\dagger} {}^j },
		
		\quad\Bl{ H_i{}W_{\rm{R}}^I{}{}^{\lambda \mu }W_{\rm{L}}^I{}{}_{\lambda }{}{}^{\nu } D_{\nu } D_{\mu } H^{\dagger} {}^i },
		
		\\& (\tau {}^K)_j^i\epsilon {}^{IJK}H^{\dagger} {}^jW_{\rm{R}}^J{}{}^{\lambda \mu }W_{\rm{L}}^I{}{}_{\lambda }{}{}^{\nu } D_{\nu } D_{\mu } H_i{} ,
		
		\quad (\tau {}^K)_j^i\epsilon {}^{IJK}W_{\rm{R}}^J{}{}^{\lambda \mu }W_{\rm{L}}^I{}{}_{\lambda }{}{}^{\nu } D_{\mu } H_i{}  D_{\nu } H^{\dagger} {}^j ,
		
		\\& H^{\dagger} {}^iW_{\rm{R}}^I{}{}^{\lambda \mu }W_{\rm{L}}^I{}{}_{\lambda }{}{}^{\nu } D_{\nu } D_{\mu } H_i{} ,
		
		\quad W_{\rm{R}}^I{}{}^{\lambda \mu }W_{\rm{L}}^I{}{}_{\lambda }{}{}^{\nu } D_{\mu } H_i{}  D_{\nu } H^{\dagger} {}^i 
		
	\end{array}\\
	
	&\begin{array}{c|l}
		
		\multirow{2}*{$\mathcal{O}_{G_{\rm{L}} H H^{\dagger}{}  G_{\rm{R}}D^2}^{\left(1\sim3\right)}$}
		
		&\Bl{ H_i{}G_{\rm{R}}^A{}{}^{\lambda \mu }G_{\rm{L}}^A{}{}_{\lambda }{}{}^{\nu } D_{\nu } D_{\mu } H^{\dagger} {}^i },
		
		\quad H^{\dagger} {}^iG_{\rm{R}}^A{}{}^{\lambda \mu }G_{\rm{L}}^A{}{}_{\lambda }{}{}^{\nu } D_{\nu } D_{\mu } H_i{} ,
		
		\\& G_{\rm{R}}^A{}{}^{\lambda \mu }G_{\rm{L}}^A{}{}_{\lambda }{}{}^{\nu } D_{\mu } H_i{}  D_{\nu } H^{\dagger} {}^i 
		
	\end{array}\label{op:flfrh2d2l}
	
\end{align}

\underline{Class $F_{\rm{R}}^2 \phi ^2 D^2$}: 4 types
\begin{align}\label{op:fr2h2d2f}
	
	&\begin{array}{c|l}
		
		\multirow{2}*{$\mathcal{O}_{H H^{\dagger}{}  B_{\rm{R}}^2D^2}^{\left(1\sim3\right)}$}
		
		&\Bl{ H_i{}B_{\rm{R}}{}{}{}^{\nu \lambda } D_{\mu } H^{\dagger} {}^i  D^{\mu } B_{\rm{R}}{}{}{}_{\nu \lambda }{} },
		
		\quad H_i{}B_{\rm{R}}{}{}_{\nu \lambda }{}B_{\rm{R}}{}{}{}^{\nu \lambda } D^{\mu } D_{\mu } H^{\dagger} {}^i ,
		
		\\& H_i{}H^{\dagger} {}^i D_{\nu } B_{\rm{R}}{}{}{}^{\lambda \mu }  D_{\mu } B_{\rm{R}}{}{}{}_{\lambda }{}{}^{\nu } 
		
	\end{array}\\
	
	&\begin{array}{c|l}
		
		\multirow{2}*{$\mathcal{O}_{H H^{\dagger}{}  B_{\rm{R}} W_{\rm{R}}D^2}^{\left(1\sim4\right)}$}
		
		&\Bl{ (\tau {}^I)_j^iH_i{}W_{\rm{R}}^I{}{}^{\nu \lambda } D_{\mu } H^{\dagger} {}^j  D^{\mu } B_{\rm{R}}{}{}{}_{\nu \lambda }{} },
		
		\quad\Bl{ (\tau {}^I)_j^iH_i{}B_{\rm{R}}{}{}_{\nu \lambda }{} D_{\mu } H^{\dagger} {}^j  D^{\mu } W_{\rm{R}}^I{}{}^{\nu \lambda } },
		
		\\& (\tau {}^I)_j^iH_i{}B_{\rm{R}}{}{}_{\nu \lambda }{}W_{\rm{R}}^I{}{}^{\nu \lambda } D^{\mu } D_{\mu } H^{\dagger} {}^j ,
		
		\quad (\tau {}^I)_j^iH_i{}H^{\dagger} {}^j D_{\mu } B_{\rm{R}}{}{}{}_{\lambda }{}{}^{\nu }  D_{\nu } W_{\rm{R}}^I{}{}^{\lambda \mu } 
		
	\end{array}\\
	
	&\begin{array}{c|l}
		
		\multirow{2}*{$\mathcal{O}_{H H^{\dagger}{}  W_{\rm{R}}^2D^2}^{\left(1\sim4\right)}$}
		
		&\Bl{ H_i{}W_{\rm{R}}^I{}{}^{\nu \lambda } D_{\mu } H^{\dagger} {}^i  D^{\mu } W_{\rm{R}}^I{}{}_{\nu \lambda }{} },
		
		\quad\Bl{ (\tau {}^K)_j^i\epsilon {}^{IJK}H_i{}W_{\rm{R}}^J{}{}^{\nu \lambda } D_{\mu } H^{\dagger} {}^j  D^{\mu } W_{\rm{R}}^I{}{}_{\nu \lambda }{} },
		
		\\& H_i{}W_{\rm{R}}^I{}_{\nu \lambda }{}W_{\rm{R}}^I{}{}^{\nu \lambda } D^{\mu } D_{\mu } H^{\dagger} {}^i ,
		
		\quad H_i{}H^{\dagger} {}^i D_{\mu } W_{\rm{R}}^I{}{}_{\lambda }{}{}^{\nu }  D_{\nu } W_{\rm{R}}^I{}{}^{\lambda \mu } 
		
	\end{array}\\
	
	&\begin{array}{c|l}
		
		\multirow{2}*{$\mathcal{O}_{H H^{\dagger}{}  G_{\rm{R}}^2D^2}^{\left(1\sim3\right)}$}
		
		&\Bl{ H_i{}G_{\rm{R}}^A{}{}^{\nu \lambda } D_{\mu } H^{\dagger} {}^i  D^{\mu } G_{\rm{R}}^A{}{}_{\nu \lambda }{} },
		
		\quad H_i{}G_{\rm{R}}^A{}_{\nu \lambda }{}G_{\rm{R}}^A{}{}^{\nu \lambda } D^{\mu } D_{\mu } H^{\dagger} {}^i ,
		
		\\& H_i{}H^{\dagger} {}^i D_{\nu } G_{\rm{R}}^A{}{}^{\lambda \mu }  D_{\mu } G_{\rm{R}}^A{}{}_{\lambda }{}{}^{\nu } 
		
	\end{array}\label{op:fr2h2d2l}
	
\end{align}

\underline{Class $\Bl{F_{\rm{L}}^2 \phi ^4 }$}: 4 types
\begin{align}\label{op:fl2h4f}
	
	&\begin{array}{c|l}
		
		\multirow{1}*{$\Bl{\mathcal{O}_{B_{\rm{L}}^2 H^2 H^{\dagger}{} ^2}}$}
		
		&\Bl{ H_i{}H_j{}H^{\dagger} {}^iH^{\dagger} {}^jB_{\rm{L}}{}{}{}_{\mu \nu }{}B_{\rm{L}}{}{}{}^{\mu \nu }}
		
	\end{array}\\
	
	&\begin{array}{c|l}
		
		\multirow{1}*{$\Bl{\mathcal{O}_{B_{\rm{L}} W_{\rm{L}} H^2 H^{\dagger}{} ^2}}$}
		
		&\Bl{ (\tau {}^I)_k^jH_i{}H_j{}H^{\dagger} {}^iH^{\dagger} {}^kB_{\rm{L}}{}{}{}_{\mu \nu }{}W_{\rm{L}}^I{}{}^{\mu \nu }}
		
	\end{array}\\
	
	&\begin{array}{c|l}
		
		\multirow{1}*{$\Bl{\mathcal{O}_{W_{\rm{L}}^2 H^2 H^{\dagger}{} ^2}^{\left(1\sim2\right)}}$}
		
		&\Bl{ (\tau {}^I)_k^i(\tau {}^J)_l^jH_i{}H_j{}H^{\dagger} {}^kH^{\dagger} {}^lW_{\rm{L}}^I{}{}_{\mu \nu }{}W_{\rm{L}}^J{}{}^{\mu \nu }},
		
		\quad\Bl{ H_i{}H_j{}H^{\dagger} {}^iH^{\dagger} {}^jW_{\rm{L}}^I{}{}_{\mu \nu }{}W_{\rm{L}}^I{}{}^{\mu \nu }}
		
	\end{array}\\
	
	&\begin{array}{c|l}
		
		\multirow{1}*{$\Bl{\mathcal{O}_{G_{\rm{L}}^2 H^2 H^{\dagger}{} ^2}}$}
		
		&\Bl{ H_i{}H_j{}H^{\dagger} {}^iH^{\dagger} {}^jG_{\rm{L}}^A{}{}_{\mu \nu }{}G_{\rm{L}}^A{}{}^{\mu \nu }}
		
	\end{array}\label{op:fl2h4l}
	
\end{align}

\underline{Class $F_{\rm{L}} \phi ^2 D^4$}: 2 types
\begin{align}\label{op:flh2d4f}
	
	&\begin{array}{c|l}
		
		\multirow{2}*{$\mathcal{O}_{B_{\rm{L}} H H^{\dagger}{} D^4}^{\left(1\sim3\right)}$}
		
		& B_{\rm{L}}{}{}{}^{\nu \lambda } D_{\nu } D^{\mu } D_{\mu } H_i{}  D_{\lambda } H^{\dagger} {}^i ,
		
		\quad B_{\rm{L}}{}{}{}^{\nu \lambda } D_{\nu } H_i{}  D^{\mu } D_{\mu } D_{\lambda } H^{\dagger} {}^i ,
		
		\\& B_{\rm{L}}{}{}{}^{\nu \lambda } D_{\mu } D_{\nu } H_i{}  D^{\mu } D_{\lambda } H^{\dagger} {}^i 
		
	\end{array}\\
	
	&\begin{array}{c|l}
		
		\multirow{2}*{$\mathcal{O}_{W_{\rm{L}} H H^{\dagger}{} D^4}^{\left(1\sim3\right)}$}
		
		& (\tau {}^I)_j^iW_{\rm{L}}^I{}{}^{\nu \lambda } D_{\nu } D^{\mu } D_{\mu } H_i{}  D_{\lambda } H^{\dagger} {}^j ,
		
		\quad (\tau {}^I)_j^iW_{\rm{L}}^I{}{}^{\nu \lambda } D_{\nu } H_i{}  D^{\mu } D_{\mu } D_{\lambda } H^{\dagger} {}^j ,
		
		\\& (\tau {}^I)_j^iW_{\rm{L}}^I{}{}^{\nu \lambda } D_{\mu } D_{\nu } H_i{}  D^{\mu } D_{\lambda } H^{\dagger} {}^j 
		
	\end{array}\label{op:flh2d4l}
	
\end{align}

\underline{Class $F_{\rm{L}} \phi ^4 D^2$}: 2 types
\begin{align}\label{op:flh4d2f}
	
	&\begin{array}{c|l}
		
		\multirow{1}*{$\mathcal{O}_{B_{\rm{L}} H^2 H^{\dagger}{} ^2D^2}^{\left(1\sim2\right)}$}
		
		&\Bl{ H_i{}H^{\dagger} {}^iB_{\rm{L}}{}{}{}^{\mu \nu } D_{\mu } H_j{}  D_{\nu } H^{\dagger} {}^j },
		
		\quad H_i{}H^{\dagger} {}^jB_{\rm{L}}{}{}{}^{\mu \nu } D_{\mu } H_j{}  D_{\nu } H^{\dagger} {}^i 
		
	\end{array}\\
	
	&\begin{array}{c|l}
		
		\multirow{3}*{$\mathcal{O}_{W_{\rm{L}} H^2 H^{\dagger}{} ^2D^2}^{\left(1\sim5\right)}$}
		
		&\Bl{ (\tau {}^I)_k^jH_i{}H_j{}W_{\rm{L}}^I{}{}^{\mu \nu } D_{\nu } H^{\dagger} {}^i  D_{\mu } H^{\dagger} {}^k },
		
		\quad\Bl{ (\tau {}^I)_k^jH_i{}H^{\dagger} {}^kW_{\rm{L}}^I{}{}^{\mu \nu } D_{\mu } H_j{}  D_{\nu } H^{\dagger} {}^i },
		
		\\& (\tau {}^I)_k^jH_i{}H^{\dagger} {}^iW_{\rm{L}}^I{}{}^{\mu \nu } D_{\mu } H_j{}  D_{\nu } H^{\dagger} {}^k ,
		
		\quad (\tau {}^I)_k^jH^{\dagger} {}^iH^{\dagger} {}^kW_{\rm{L}}^I{}{}^{\mu \nu } D_{\mu } H_i{}  D_{\nu } H_j{} ,
		
		\\& (\tau {}^I)_k^jH_j{}H^{\dagger} {}^iW_{\rm{L}}^I{}{}^{\mu \nu } D_{\mu } H_i{}  D_{\nu } H^{\dagger} {}^k 
		
	\end{array}\label{op:flh4d2l}
	
\end{align}

\underline{Class $F_{\rm{R}} \phi ^4 D^2$}: 2 types
\begin{align}\label{op:frh4d2f}
	
	&\begin{array}{c|l}
		
		\multirow{1}*{$\mathcal{O}_{H^2 H^{\dagger}{} ^2 B_{\rm{R}}D^2}$}
		
		&\Bl{ H_i{}H^{\dagger} {}^jB_{\rm{R}}{}{}{}^{\mu \nu } D_{\mu } H_j{}  D_{\nu } H^{\dagger} {}^i }
		
	\end{array}\\
	
	&\begin{array}{c|l}
		
		\multirow{1}*{$\mathcal{O}_{H^2 H^{\dagger}{} ^2 W_{\rm{R}}D^2}^{\left(1\sim2\right)}$}
		
		&\Bl{ (\tau {}^I)_l^jH_i{}H_j{}W_{\rm{R}}^I{}{}^{\mu \nu } D_{\mu } H^{\dagger} {}^i  D_{\nu } H^{\dagger} {}^l },
		
		\quad\Bl{ (\tau {}^I)_l^jH_i{}H^{\dagger} {}^lW_{\rm{R}}^I{}{}^{\mu \nu } D_{\mu } H_j{}  D_{\nu } H^{\dagger} {}^i }
		
	\end{array}\label{op:frh4d2l}
	
\end{align}

\underline{Class $\phi ^2 D^6$}: 1 type
\begin{align}\label{op:h2d6}
	
	&\begin{array}{c|l}
		
		\multirow{1}*{$\mathcal{O}_{H H^{\dagger}{} D^6}$}
		
		& H_i{} D^{\lambda } D_{\lambda } D_{\mu } D^{\mu } D^{\nu } D_{\nu } H^{\dagger} {}^i 
		
	\end{array}
	
\end{align}

\underline{Class $\phi ^4 D^4$}: 1 type
\begin{align}\label{op:h4d4}
	
	&\begin{array}{c|l}
		
		\multirow{7}*{$\mathcal{O}_{H^2 H^{\dagger}{} ^2D^4}^{\left(1\sim13\right)}$}
		
		&\Bl{ H_i{}H_j{} D_{\nu } D_{\mu } H^{\dagger} {}^i  D^{\nu } D^{\mu } H^{\dagger} {}^j },
		
		\quad\Bl{ H_i{}H^{\dagger} {}^i D_{\nu } D_{\mu } H_j{}  D^{\nu } D^{\mu } H^{\dagger} {}^j },
		
		\\&\Bl{ H_i{} D_{\mu } H_j{}  D_{\nu } H^{\dagger} {}^i  D^{\nu } D^{\mu } H^{\dagger} {}^j },
		
		\quad H_i{}H^{\dagger} {}^iH^{\dagger} {}^j D_{\mu } D^{\mu } D^{\nu } D_{\nu } H_j{} ,
		
		\\& H_i{}H_j{}H^{\dagger} {}^j D_{\mu } D^{\mu } D^{\nu } D_{\nu } H^{\dagger} {}^i ,
		
		\quad H_i{}H^{\dagger} {}^j D_{\nu } D^{\mu } D_{\mu } H_j{}  D^{\nu } H^{\dagger} {}^i ,
		
		\\& H_i{}H^{\dagger} {}^i D_{\nu } D^{\mu } D_{\mu } H_j{}  D^{\nu } H^{\dagger} {}^j ,
		
		\quad H_i{}H^{\dagger} {}^j D_{\mu } H_j{}  D^{\nu } D_{\nu } D^{\mu } H^{\dagger} {}^i ,
		
		\\& H_i{}H^{\dagger} {}^i D_{\mu } H_j{}  D^{\nu } D_{\nu } D^{\mu } H^{\dagger} {}^j ,
		
		\quad H_i{}H^{\dagger} {}^j D^{\mu } D_{\mu } H_j{}  D^{\nu } D_{\nu } H^{\dagger} {}^i ,
		
		\\& H_i{}H^{\dagger} {}^i D^{\mu } D_{\mu } H_j{}  D^{\nu } D_{\nu } H^{\dagger} {}^j ,
		
		\quad H_i{}H_j{} D^{\mu } D_{\mu } H^{\dagger} {}^i  D^{\nu } D_{\nu } H^{\dagger} {}^j ,
		
		\\& H_i{} D^{\mu } D_{\mu } H_j{}  D_{\nu } H^{\dagger} {}^i  D^{\nu } H^{\dagger} {}^j 
		
	\end{array}
	
\end{align}

\underline{Class $\phi ^6 D^2$}: 1 type
\begin{align}\label{op:h6d2}
	
	&\begin{array}{c|l}
		
		\multirow{2}*{$\mathcal{O}_{H^3 H^{\dagger}{} ^3D^2}^{\left(1\sim4\right)}$}
		
		&\Bl{ H_i{}H_j{}H_k{}H^{\dagger} {}^i D_{\mu } H^{\dagger} {}^j  D^{\mu } H^{\dagger} {}^k },
		
		\quad\Bl{ H_i{}H_j{}H^{\dagger} {}^iH^{\dagger} {}^j D_{\mu } H_k{}  D^{\mu } H^{\dagger} {}^k },
		
		\\& H_i{}H_j{}H^{\dagger} {}^iH^{\dagger} {}^k D_{\mu } H_k{}  D^{\mu } H^{\dagger} {}^j ,
		
		\quad H_i{}H_k{}H^{\dagger} {}^iH^{\dagger} {}^jH^{\dagger} {}^k D^{\mu } D_{\mu } H_j{} 
		
	\end{array}
	
\end{align}

\underline{Class $\Bl{\phi ^8 }$}: 1 type
\begin{align}\label{op:h8}
	
	&\begin{array}{c|l}
		
		\multirow{1}*{$\Bl{\mathcal{O}_{H^4 H^{\dagger}{} ^4}}$}
		
		&\Bl{ H_i{}H_{j}{}H_k{}H_{l}{}H^{\dagger} {}^iH^{\dagger} {}^{j}H^{\dagger} {}^kH^{\dagger} {}^{l}}
		
	\end{array}
	
\end{align}

\subsection{Operators Involving Fermions}\label{sec:fermionopes}
In this subsection, we list the Green's basis of operators involving fermions at the dimension 8 in the SMEFT. The fermions in the SM are presented as the two-component Weyl spinors during the construction of the basis and are converted to the four-component Dirac spinors in the final result for the reader's convenience. In the SM, the four-component Dirac spinors are related to the two-component Weyl spinors by the following formulas:
\begin{align}
q_{\rm{L}}=\begin{pmatrix}Q\\0\end{pmatrix},\quad u_{\rm{R}}=\left(\begin{array}{c}0\\u_{_\mathbb{C}}^{\dagger}\end{array}\right),\quad d_{\rm R}=\left(\begin{array}{c}0\\d_{_\mathbb{C}}^{\dagger}\end{array}\right),\quad l_{\rm L}=\left(\begin{array}{c}L\\0\end{array}\right),\quad e_{\rm R}=\left(\begin{array}{c}0\\e_{_\mathbb{C}}^{\dagger}\end{array}\right).\\
\bar{q}_{\rm{L}}=\left(0\,,\,Q^{\dagger} \right),\quad \bar{u}_{\rm{R}}=\left(u_{_\mathbb{C}}\,,\,0 \right),\quad \bar{d}_{\rm R}=\left(d_{_\mathbb{C}}\,,\,0\right),\quad \bar{l}_{\rm L}=\left(0\,,\,L^{\dagger}\right),\quad \bar{e}_{\rm R}=\left(e_{_\mathbb{C}}\,,\,0\right).
\end{align}
In the following, we will omit the $\rm{L/R}$ in the subscript of each four-component fermion. The Hermitian conjugates of four-component spinor bilinears can be converted by the following relations:
\eq{
	\left(\bar{\Psi}_1\Psi_2\right)^\dagger &=\bar{\Psi}_2\Psi_1,\\
	\left(\bar{\Psi}_1\gamma^{\mu}\Psi_2\right)^\dagger &=\bar{\Psi}_2\gamma^{\mu}\Psi_1,\\
	\left(\bar{\Psi}_1\sigma^{\mu\nu}\Psi_2\right)^\dagger &=\bar{\Psi}_2\sigma^{\mu\nu}\Psi_1,\\
	\left(\Psi^{\rm{T}}_1C\Psi_2\right)^\dagger &=\bar{\Psi}_2C\bar{\Psi}^{\rm{T}}_1,\\
	\left(\Psi^{\rm{T}}_1C\gamma^{\mu}\Psi_2\right)^\dagger &=\bar{\Psi}_2\gamma^{\mu}C\bar{\Psi}_1^{\rm{T}},\\
	\left(\Psi^{\rm{T}}_1C\sigma^{\mu\nu}\Psi_2\right)^\dagger &=\bar{\Psi}_2\sigma^{\mu\nu}C\bar{\Psi}_1^{\rm{T}},
} 
where in the chiral representation $C=i\gamma^0\gamma^2=\begin{pmatrix} \epsilon_{\alpha\beta}&0\\0&\epsilon^{\dot{\alpha}\dot{\beta}}\end{pmatrix}=\begin{pmatrix} -\epsilon^{\alpha\beta}&0\\0&-\epsilon_{\dot{\alpha}\dot{\beta}}\end{pmatrix}$,  $\gamma^{\mu}=\begin{pmatrix}
0&\sigma^{\mu}_{\alpha\dot{\beta}}\\\bar{\sigma}^{\mu\dot{\alpha}\beta}&0
\end{pmatrix}$ and $\sigma^{\mu\nu}=\dfrac{i}{2}[\gamma^\mu,\gamma^\nu]=\begin{pmatrix}
\left(\sigma^{\mu\nu}\right)_\alpha{}^\beta&0\\0&\left(\bar{\sigma}^{\mu\nu}\right)^{\dot{\alpha}}{}_{\dot{\beta}}
\end{pmatrix}$.
\\ \\

\input{Dim_8_Fermionic_Operators}

\section{Conclusion}\label{sec:con}

In this paper, we proposed an off-shell amplitude formalism to construct any general operators in term of the spinor formalism and then remove various kinds of redundancies among effective operators step-by-step. In this formalism, any redundant operators that can relate to each other by all redundancies correspond to off-shell amplitudes via a one-to-one mapping. We define the off-shell spinors, in which the spinor indices from the covariant derivatives and fields are different and thus distinguished, so that the off-shell amplitudes encode the information of the EOM and the CDC. We utilized the formalism to construct the Green's basis in the SMEFT in a systematic way, with the following algorithm:
\bit
\item Firstly an over-complete off-shell Lorentz basis of operators, carrying all kinds of redundancies, is built from all the contractions of possible off-shell spinors.
\item All the redundancy relations such as the IBP, the Schouten identity, the CDC and the Bianchi identity, are written in the off-shell amplitude formalism. 
\item Utilizing these relations, the over-complete set of operators can be reduced systematically to the ones with only the EOM kept. 
\item Finally, with the method introduced and developed in Ref.~\cite{Li:2020gnx,Li:2020xlh,Li:2022tec}, the gauge basis and the flavor basis are constructed to obtain the complete and independent off-shell operator basis, as the Green's basis. 
\eit

The complete statistic of the basis is given in table~\ref{tab:statistic1} and table~\ref{tab:statistic2}. According to our result, we conclude that there are 993+1649=2642 independent operators for one generation of fermions and 44807+66197=111004 independent operators for three generation of fermions, which is consistent with the counting result given by the Hilbert Series and Sym2Int. The method can be applied to general EFTs at any mass dimension.

An advantage of the off-shell amplitude formalism introduced in this paper is that it completely preserve the off-shell information of operators when mapping operators to off-shell amplitudes. Thus the off-shell amplitude formalism provide a systematic way to reduce the Green's basis into an complete and independent on-shell physical basis with the EOM and the CDC. Furthermore, it is possible to reduce any operator or operator basis into an complete and independent on-shell physical basis and achieve conversions among different operator bases systematically with the off-shell amplitude formalism.

\section*{Acknowledgments}

We thank Ming-Lei Xiao, Hao-Lin Li and Yi-Ning Wang for valuable discussions, and Jose Santiago for valuable suggestions and communications. 
Z. R. and J.-H.Y. are supported by the National Science Foundation of China under Grants No. 12022514, No. 11875003 and No. 12047503, and National Key Research and Development Program of China Grant No. 2020YFC2201501, No. 2021YFA0718304, and CAS Project for Young Scientists in Basic Research YSBR-006, the Key Research Program of the CAS Grant No. XDPB15.

\bibliographystyle{JHEP}
\bibliography{MyRef}

\end{document}